\documentclass[aps,prb,reprint,superscriptaddress,floatfix]{revtex4-2}

\usepackage{graphicx}
\usepackage{tikz}
\usepackage{amsmath}
\usepackage{amssymb}

\begin{document}


\title{Dyakonov-like surface waves in anisotropic cylindrical waveguides}


\author{K.~Yu. Golenitskii}
\email[]{golenitski.k@mail.ioffe.ru}
\affiliation{Ioffe Institute, 194021 St.~Petersburg, Russia}

\author{A.~A. Bogdanov}
\affiliation{Ioffe Institute, 194021 St.~Petersburg, Russia}
\affiliation{ITMO University, 197101 St.~Petersburg, Russia}


\date{\today}

\begin{abstract}
Dyakonov surface waves are a well-known example of surface electromagnetic waves propagating along a plane interface between isotropic and anisotropic dielectric media. Here, we investigate the spectrum and radiative losses of Dyakonov waves at the curved interface considering both cases of \textquotedblleft{}positive\textquotedblright{} and \textquotedblleft{}negative\textquotedblright{} curvature of anisotropic medium. We demonstrate how Dyakonov waves at a plane interface continuously transform to the leaky or guided modes of an anisotropic cylindrical waveguide and simple derive asymptotic equation for radiative losses.  All analytical results are confirmed by solving exact dispersion equation numerically. We believe that our work extends the potential practical application of Dyakonov waves in optics and photonics devices.





\end{abstract}


\maketitle

\section{Introduction}
Dyakonov surface electromagnetic waves (DSWs) attract much attention in recent years \cite{Takayama2008DyakonovReview,Takayama2017PhotonicInterfaces} because they can propagate along a plane interface of an anisotropic dielectric medium.
Thus, they have vanishing losses due to absorption that opens rich prospective for practical applications.
DSWs were predicted by Michel Dyakonov in 1988~\cite{Dyakonov1988} for the interface between an isotropic dielectric and uniaxial dielectric crystal.
Further, existence of DSWs was demonstrated for many kinds of partnering media including uniaxial-uniaxial \cite{Averkiev1990}, isotropic-biaxial \cite{Walker1998SurfaceInterfaces,Cojocaru2015DyakonovInterface,Jacob2008OpticalMetamaterials}, biaxial-biaxial \cite{JohnA.Polo2007SurfaceInterface}, chiral media \cite{Agarwal_2009,Gao2010Dyakonov-TammHandedness}, metamaterials \cite{Jacob2008OpticalMetamaterials,Takayama2017PhotonicInterfaces}.


In most of the cases, the propagation of DSWs along a flat boundary are considered.
In this work, we aim at studying the spectrum, radiative losses and mode profiles of DSWs at a curved interface.
A similar analysis has been done for some types of surface waves.
For example, the propagation of surface plasmon polaritons (SPPs) along curved plasmonic structures is considered in works \cite{Kim2006LeakyWaveguide,Kotelnikov2015ElectromagneticCylinder,Fang2015NanoplasmonicCircuits}, where it was shown that SPP can turn into either guided or leaky mode of the metallic wire depending on the propagation direction of SPP with respect to the wire axis.
We show that DSWs mostly turn into the leaky modes of anisotropic waveguide and they turn into the guided modes only for very specific materials and geometrical parameters of the system.

It is convenient to consider a curved interface as a cylindrical waveguide, which is well-studied for both cases of metal and dielectric core/cladding.
Usually, the materials of the waveguide are considered as isotropic.
Dielectric anisotropic waveguides were also considered in Ref.~\cite{Snyder1984OpticalTheory} without referring to the surface modes.
It is necessary to say that the weak anisotropy does not bring qualitative changes of the waveguide modes until conditions for DSWs are met.
To the best of our knowledge, there is no works on DSWs or their analogues in a conventional  step-index anisotropic cylindrical waveguides.
In the paper \cite{Kajorndejnukul2019ConformalMetamaterialsb} authors are using transformation optics technique to map DSWs into bound waveguide modes.
It results in using gradient-index metamaterial waveguide but not traditional step-index waveguides.

The eigenmodes of the lossless waveguide (without material absorption) can be classified in two types: (i) guided modes without radiative losses and leaky modes with radiative losses to the surrounding space~\cite{Snyder1984OpticalTheory}.
It is not entirely clear what should we call surface modes in this case.
A surface wave can be both guided as SPP under the light cone or leaky (resonant) as a SPP above the light cone.
There is no canonical definition of surface waves in waveguides but we can define it as waves (leaky or guided) transforming to true surface waves as the radius of the cylinder tends to infinity.
The definition of surface waves in a cylinder of smaller radii is not so evident and it will be discussed below in Sec. \ref{sec:arbitrary-waveguide}.A.

A conventional waveguide modes (WMs) are the example of guided modes \cite{Snyder1984OpticalTheory}.
Whispering gallery modes (WGMs) in a dielectric bottle resonators \cite{Sumetsky2004Whispering-gallery-bottleEtalonb} are the example of leaky modes (see review in Ref. \cite{Oraevsky2002Whispering-galleryWaves}).
Both WMs and WGMs are not truly surface modes because they are mainly localized inside, not on the interface of the waveguide.
For surface wave we expect the maximum field intensity at the boundary of the waveguide but not inside its core as it happens for WMs and WGMs.
The existence of DSWs in a cylindrical anisotropic waveguide of a large radii is almost obvious.
Indeed, the curvature can be considered as a perturbation, which affect the mode profile only in the next order of the perturbation theory after periodicity condition.
For smaller radii, the situation is not so evident as the field profile demonstrate more complicated behaviour, see Sec.~\ref{sec:arbitrary-waveguide}.D.


The paper is organized as follows.
In Sec.~\ref{sec:model} we describe the model of anisotropic waveguide and dispersion equation.
Section~\ref{sec:large-waveguide} is devoted to an analysis of modes in large radius waveguides which can be considered almost flat.
Arbitrary radius waveguides are analyzed in Sec.~\ref{sec:arbitrary-waveguide}.
The radiation losses of Dyakonov-like modes are considered in Sec.~\ref{sec:losses}.
The results are summarized in Sec.~\ref{sec:conclusion}.

\section{Model}\label{sec:model}
\begin{figure}
    \centering
    \begin{tikzpicture}
    \node (image1) at (0,0) {\includegraphics[page=1,height=4.5cm,clip]{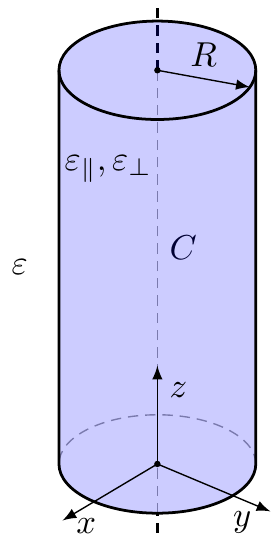}};
    \node (image2) at (4.5,0) {\includegraphics[page=2,height=4.5cm,clip]{sketch.pdf}};
    \node[below] at (image1.south) {(a) Case A};
    \node[below] at (image2.south) {(b) Case B};
    \end{tikzpicture}
    \caption{Geometry of the cylindrical waveguide in a dielectric medium. (a) the anisotropic waveguide in an isotropic medium. (b) the isotropic waveguide in an anisotropic medium.}
    \label{fig:sketch}
\end{figure}

In the original work \cite{Dyakonov1988}, DSW propagates along the interface between an isotropic dielectric and anisotropic uniaxial dielectric crystal.
There are two ways to bend the interface one is in the direction of the anisotropic medium, the other, in the direction of the isotropic medium.
For simplicity, let us consider an infinite cylindrical waveguide of a radius \(R\) placed inside the other dielectric medium.
Thus, there are two structures of the waveguide possible following the interface bending.
The first is waveguide made of anisotropic medium and surrounded by isotropic dielectric medium [Fig.~\ref{fig:sketch}(a)], which we also call as \textquotedblleft{}positive\textquotedblright{} curvature.
The second is waveguide made of isotropic medium and surrounded by anisotropic dielectric medium [Fig.~\ref{fig:sketch}(b)], which we also call as \textquotedblleft{}negative\textquotedblright{} curvature.
Further in the text, we will refer to these cases as Case A and Case B, respectively.
Moreover, to simplify the analysis, the optical axis of the uniaxial material is parallel to the waveguide axis in both cases.
Following Dyakonov \cite{Dyakonov1988}, we consider positive uniaxial material with dielectric permittivity tensor $\hat{\varepsilon} = \operatorname{diag}(\varepsilon_\perp, \varepsilon_\perp, \varepsilon_\|)$ and isotropic medium with dielectric permittivity $\varepsilon$.
$\varepsilon_\|$ corresponds to $z$ axis in this geometry, while $\varepsilon_\perp$ corresponds to $(x,y)$ or $(\rho,\varphi)$ plane perpendicular to the optical axis.
Also, for DSWs propagation it is necessary to fulfill the condition $\varepsilon_\perp < \varepsilon < \varepsilon_\|$ at some frequency.

\subsection{Dispersion equations}
Let us give a brief reminder of the derivation of the dispersion equation $\omega(q)$ in the case of cylindrical waveguides \cite{Snyder1984OpticalTheory}.
One can seek the solutions of Maxwell's equations in the form of monochromatic waves traveling along the $z$ direction $\mathbf{E}, \mathbf{H} \propto \operatorname{exp}(i q z - i \omega t)$.
Since $\hat{\varepsilon}$ is diagonal in cylindrical coordinates $(\rho,\varphi,z)$ Maxwell's equations can be easily separated in two modes corresponded to $E_z$ and $H_z$, respectively.
These modes are also called as E- and H-wave, are independent until the boundary conditions mix them up.
The following equations describe field components distribution:
\begin{align}
    \Delta E_z + k_e^2 E_z &= 0, \label{eq:ez}\\
    \Delta H_z + k_o^2 H_z &= 0, \label{eq:hz}
\end{align}
where $\Delta$ is 2D Laplace operator in polar coordinates, $k_e^2 = \varepsilon_\| k_o^2 / \varepsilon_\perp$ and $k_o^2 = \varepsilon_\perp \omega^2/c^2 - q_z^2$, and $c$ is speed of light.
For the waves in the isotropic medium one should replace $\varepsilon_\| = \varepsilon_\perp = \varepsilon$ from which it immediately follows $k_e^2 = k_o^2 \equiv k^2$.
Expressions for other field components in terms of $E_z$ and $H_z$ are given in Appendix \ref{appendix:fields}.
The separation of variables $E_z, H_z \propto \mathrm{f}_\nu(\rho)\exp(i\nu \varphi)$ in Eq. \eqref{eq:ez} and Eq. \eqref{eq:hz} leads to Bessel equation
\begin{equation}
    \frac{1}{\rho}\frac{\mathrm{d}}{\mathrm{d}\rho}\left(\rho \frac{\mathrm{d}\mathrm{f}_\nu}{\mathrm{d}\rho}\right) + \left( k^2 - \frac{\nu^2}{\rho^2} \right) \mathrm{f}_\nu = 0, \label{eq:bessel_equation}
\end{equation}
\begin{table}[b]
\caption{\label{tab:solutions}Appropriate solutions $f_\nu$ in cylindrical waveguide}
\begin{ruledtabular}
\begin{tabular}{c|cc}
 & $\rho < R$ & $\rho > R$\\\hline
$\quad k^2 > 0, \varkappa^2 < 0 \quad$ & $J_\nu (k\rho)$ & $Y_\nu(k\rho)$ or $H_\nu^{(1)}(k\rho)$ \\
$\quad k^2 < 0, \varkappa^2 > 0 \quad$ & $I_\nu (\varkappa\rho)$ & $K_\nu(\varkappa\rho)$ \\
\end{tabular}
\end{ruledtabular}
\end{table}
where $k^2 = k_e^2$ for E-wave and $k^2 = k_o^2$ for H-wave.

The sign of $k^2$ varies with $\omega$ and $q$.
Here and what follows we use $k$ in formulae implying that $k^2 > 0$.
In other case $k^2 < 0$, we use $\varkappa$ defined as $\varkappa^2 = -k^2 > 0$.
Depending on the region $\rho < R$ or $\rho > R$ one should take appropriate solution of Eq. \eqref{eq:bessel_equation}.
$f_\nu$ should be finite at $\rho = 0$ in the medium $\rho < R$, then $f_\nu = J_\nu(k\rho), I_\nu(\varkappa \rho)$.
The choice of $f_\nu$ in the region $\rho > R$ depends on the statement of the problem.
Square-integrable eigenmodes on $\rho > R$ exist if and only if $k^2 < 0$ and then $f_\nu = K_\nu(\varkappa \rho)$.
If $k^2 > 0$ then one could search for generalized eigenmodes or leaky modes.
For the problem of generalized eigenmodes the appropriate solution is $f_\nu = Y_\nu(k \rho)$ that yields solutions for $q_z$ in reals.
In the case of leaky modes one should take $f_\nu = H^{(1)}_\nu(k\rho)$.
It is diverging cylindrical wave cause $H^{(1)}_\nu(k\rho) \sim \exp(ik\rho)/\sqrt{\rho}$ as $k \rho \xrightarrow{} \infty$.
Using this $f_\nu = H^{(1)}_\nu(k\rho)$ leads to complex-valued $q_z = \operatorname{Re} q_z + i \operatorname{Im} q_z$ and $\operatorname{Im} q_z < 0$ that corresponds to the damping of the mode along the waveguide.
Table \ref{tab:solutions} summarizes the choosing of solutions.

After, one should satisfy Maxwell's boundary conditions on the interface between media and get the dispersion equation.
The following dispersion equation is obtained for waveguide in Case A.
\begin{equation}
	C_h C_e = \frac{\nu^2 q_z^2 \omega^2}{k_o^4 k^4 c^2 R^2} (\varepsilon - \varepsilon_\perp)^2, \label{eq:main}
\end{equation}
where
\begin{align}
    C_h &= \frac{F^{\text{in}}_h}{k_o^2} - \frac{F^{\text{out}}_h}{k^2}, \\
    C_e &= \frac{\varepsilon_\perp F^{\text{in}}_e}{k_o^2} - \frac{\varepsilon F^{\text{out}}_e}{k^2},
\end{align}
and $F^{\text{in},\text{out}}_{h,e} = \frac{\mathrm{d}}{\mathrm{d} \rho} \ln \mathrm{f}_\nu(\rho)\big|_{\rho=R}$, where $\mathrm{f}_\nu(\rho)$ is appropriate solution of Eq. \eqref{eq:bessel_equation}, indexes $e$ and $h$ correspond to E-wave and H-wave, superscripts ${}^\text{in}$ and ${}^\text{out}$ correspond to the region $\rho < R$ and $\rho > R$, respectively.
See Appendix \ref{appendix:main_equation} for more details.
To obtain dispersion equation in Case B one should swap $k_o^2$ and $k^2$, $\varepsilon_\perp$ and $\varepsilon$, and choose another $\mathrm{f}_\nu(\rho)$ if needed.

For simplicity of the analysis, we neglect the frequency dispersion of $\varepsilon(\omega), \varepsilon_\perp(\omega), \varepsilon_\|(\omega)$ further in the text, since it does not affect the qualitative result.
In what follows all dimensional values such as $\mathbf{q}, R$ are measured in units of $\omega/c$ unless otherwise stated.

\section{Eigenmodes in a large radius $R$ waveguides}\label{sec:large-waveguide}
\subsection{\textquotedblleft{}Quantization\textquotedblright{} of Dyakonov waves}
DSWs in a waveguide should not be much different from DSWs on a flat interface in the case of waveguides of a large radius $R \gg \lambda$, where $\lambda$ is wavelength.
The more precise ratio between $R$ and $\lambda$ is related to a narrow range of DSWs propagation angles $\theta$ along a flat boundary.

\begin{figure}
    \centering
    \includegraphics[width=\linewidth,clip,page=2]{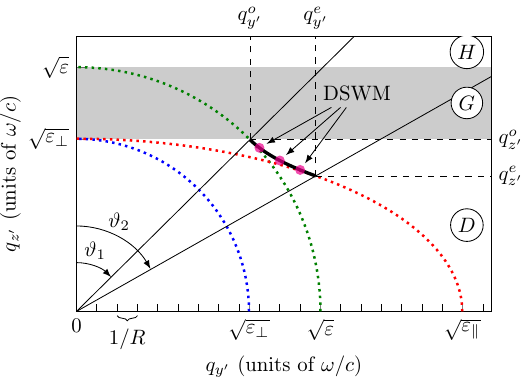}
    \caption{Dyakonov surface wave (DSW) angular dispersion (black solid line) on a flat interface between isotropic medium $\varepsilon = 4$ and uniaxial medium $\varepsilon_\| = 10, \varepsilon_\perp = 2$. The optical axis coincides with the $z'$ axis of the waveguide. $y'=\varphi/R, z'=z$ are local coordinates at the waveguide boundary. Blue dotted circle shows the dispersion of TE- and TM-plane wave in the isotropic medium. Red and green dotted curves show the dispersion of extraordinary and ordinary plane waves in uniaxial medium. $\vartheta_1, \vartheta_2$ show the limit of allowable propagation angles $\vartheta$ for DSWs. $q_{z'}^o$ and $q_{z'}^e$ correspond to the range of possible values of $q_{z'}$ for DSWs, respectively. Vertical ticks show discrete values $q_{y'} = \nu/R, \nu \in \mathbb{Z}$ for $R=7 \cdot c/\omega$. $D,G,H$ mark specific ranges of $q_z = q_{z'}$ in the waveguide corresponded to leaky or guided modes. In Case A [Fig. \ref{fig:sketch}(a)] modes with $q_z$ from regions $D$ and $G$ are leaky, and from region $H$ are guided. In Case B [Fig. \ref{fig:sketch}(b)] modes with $q_z$ from region $D$ are leaky and from regions $G$ and $H$ are guided. The purple circles on the DSWs dispersion curve mark approximate solutions $(q_z,\nu/R)$ for Dyakonov-like surface waveguide modes in the cylindrical waveguide [Fig.~\ref{fig:sketch}] with $R=7 \cdot c/\omega$ and azimuthal numbers $\nu=9, 10, 11$.}
    \label{fig:flat_surface}
\end{figure}

Firstly, let us consider DSW propagating along the flat interface between isotropic medium $\varepsilon = 4$ and anisotropic uniaxial medium $\varepsilon_\| = 10, \varepsilon_\perp = 2$, as an example.
$(y',z')$ are Cartesian coordinates on the interface with which we will then associate the local coordinates on the curved waveguide surface.
$z'$-axis coincides with the optical axis of the uniaxial medium.
The black curve in Fig.~\ref{fig:flat_surface} shows the DSWs angular dispersion of wave-vector $\mathbf{q}=(q_{y'},q_{z'})$.
DSWs can only propagate in a small range $(\vartheta_1, \vartheta_2)$ of angles $\vartheta$ relative to the optical axis \cite{Dyakonov1988}.
The dotted circles and ellipse show the angular dispersion of plane waves in media with wave-vectors parallel to the interface.
Green dotted curve is TE- and TM-polarized plane waves in the isotropic medium
\[
    q_{z'}^2 + q_{y'}^2 = \varepsilon.
\]
Blue and red dotted curves are ordinary and extraordinary waves in the uniaxial medium
\begin{align}
        q_{z'}^2 + q_{y'}^2 &= \varepsilon_\perp, \\
        \frac{q_{y'}^2}{\varepsilon_\|} + \frac{q_{z'}^2}{\varepsilon_\perp} &= 1.\label{eq:extraordinary-wave}
\end{align}
The dispersion curve of surface waves, including DSWs too, lies under the line cone of ordinary and extraordinary waves.

Two effects of the curvature of a cylindrical waveguide are quantization of $q_{y'}$ and possible occurrence of radiation losses for some modes.
Let us define local Cartesian coordinates on the waveguide surface as $z' = z, y' = \varphi R$.
Polar angle $\varphi \in [0; 2\pi)$ is a cyclic variable.
Then $y'$ is a cyclic variable too, and it leads to the cyclic boundary conditions $\mathbf{E}(y') = \mathbf{E}(y' + 2\pi R)$ and $\mathbf{H}(y') = \mathbf{H}(y' + 2\pi R)$.
It means that wave-vector component $q_{y'}$ corresponded to $y'$ can take a discrete set of values $q_{y',\nu} = \nu / R$ where $\nu \in \mathbb{Z}$.
The possible values of $q_{y',\nu}$ in the waveguide of radius $R = 7$ are marked by ticks in Fig.~\ref{fig:flat_surface}.
Thereby, there are three points $(q_{z'},q_{y',\nu})$ with $\nu=9, 10, 11$ falling on the DSWs dispersion curve.
These points correspond to eigenmodes in the waveguide.
We will refer to them as Dyakonov-like surface waveguide mode (DSWM).
In Fig.~\ref{fig:flat_surface} DSWMs are marked by purple dots on the DSWs dispersion black curve.

The total number of DSWMs depends on the range of allowable values $D_{y'} = (q_{y'}^o, q_{y'}^e)$ of $q_{y'}$ for the DSWs on a flat interface [Fig.~\ref{fig:flat_surface}] and the radius of waveguide $R$.
The range strongly depends on anisotropy factor ${\eta = \frac{\varepsilon_\|}{\varepsilon_\perp} - 1 > 0}$ and $\xi = \frac{\varepsilon - \varepsilon_\perp}{\varepsilon_\|-\varepsilon_\perp} \in (0; 1)$.
It is clear, that the step of quantization $\Delta q_{y'} = 1/R$ increases as $R \rightarrow 0$ and $\Delta q_{y'} > \sqrt{\varepsilon_\perp}, \sqrt{\varepsilon}, \sqrt{\varepsilon_\|}$ at some moment.
It is the limit of narrow waveguides $R \lesssim \lambda$.
On the other hand, DSWs have a limited region of allowed values $q_{y'}$.
There are waveguide radii $R$ such that not a single value of $q_{y',\nu}$ falls into the DSWs range $D_{y'}$.
Let us define the minimal radius $R_0$ that at least one $q_{y',\nu}$ falls into $D_{y'}$ for any $R > R_0$.
It is clear that if $\Delta q_{y'} < q_{y'}^e - q_{y'}^o$ then such $q_{y',\nu}$ exists.
So $R_0$ is defined by corner case $\Delta q_{y'} = 1/R_0 = q_{y'}^e - q_{y'}^o$.
The condition $R_0 > \lambda$ is always fulfilled until we consider exotic cases of extremely large $\eta$.

Using the definition of DSWs angular range given in Dyakonov's paper \cite{Dyakonov1988} one can get $q_{y'}^o, q_{y'}^e$ and then evaluate $R_0(\eta, \xi)$ numerically.
In the limit of weak anisotropy ${\eta \ll 1}$, analytical expression for $R_0$ can be obtained
\begin{gather}
    \frac{2\pi R_0}{\lambda_\varepsilon} = \frac{1}{\eta^2 \sqrt{\xi} (1-\xi)^2}, \label{eq:r0_weak}
\end{gather}
where $\lambda_\varepsilon = \lambda_0 / \sqrt{\varepsilon}$ is wavelength in isotropic medium, $\lambda_0$ is wavelength in vacuum.
See Appendix \ref{appendix:minimal_radius} for a detailed derivation of Eq. \eqref{eq:r0_weak} and the case of strong anisotropy $\eta > 1$.
It can be seen from Eq. \eqref{eq:r0_weak} that $R_0/\lambda_\varepsilon \xrightarrow{} \infty$ as $\xi \xrightarrow{} 0$ or $\xi \xrightarrow{} 1$ that corresponds to the limits $\varepsilon \xrightarrow{} \varepsilon_\perp$ or $\varepsilon \xrightarrow{} \varepsilon_\|$.
This is because the width of the allowable range of angles $\varphi$ for DSWs tends to 0 in these limits.
For each fixed anisotropy factor $\eta$ there is the optimal value $\xi^*$ when $R_0$ is minimal.
$\xi^* = 1/5$ in the limit of weak anisotropy $\eta \ll 1$ [Eq. \eqref{eq:r0_weak}].

\subsection{Radiation leakage}
The next step is taking into account the possibility of radiation losses due to the bending of the interface in the waveguide.
As mentioned before, all modes in the cylindrical waveguide are leaky or guided \cite{Snyder1984OpticalTheory}.
Guided modes do not have radiation leakage by definition.
In other words they are propagating losslessly if we neglect media losses $\operatorname{Im} \varepsilon = \operatorname{Im} \hat{\varepsilon} = 0$.
According to Eqs. \eqref{eq:ez} and \eqref{eq:hz} modes are guided if both $k_e^2, k_o^2 < 0$ for the medium outside.
In other cases, modes are leaky.
$\varepsilon_\perp, \varepsilon_\|, \varepsilon > 0$ in our model.
Then, inequalities are reduced to $q_z^2 > \varepsilon$ in Case A and to $q_z^2 > \varepsilon_\perp$ in Case B.
The corresponding areas of $q_{z}=q_{z'}$ for both cases are marked with $D$, $G$, and $H$ in Fig.~\ref{fig:flat_surface}.

In Case A [Fig.~\ref{fig:sketch}(a)] leaky modes are lying in areas $D$ and $G$ and guided modes are lying in area $H$.
Since $\varepsilon_\perp < \varepsilon$ then it is known that there are no guided modes in area $H$.
A waveguide in Case B [Fig.~\ref{fig:sketch}(b)] supports guided modes and they are lying in areas $G$ and $H$ in Fig.~\ref{fig:flat_surface}.
As in Case A, leaky modes in Case B are lying in area~$D$.
The dispersion curve of DSWs from the example [Fig.~\ref{fig:flat_surface}] completely falls into area $D$, $q_z^2 < \varepsilon_\perp$.
Then DSWMs should be leaky in both types of waveguide but this is not true for any $\eta, \xi$.

The more accurate analysis shows that there is a specific case of parameters when some DSWMs could be guided.
If the DSWs dispersion curve partially falls into region $q_z^2 > \varepsilon_\perp$ then the existence of guided DSWMs is possible in Case B.
In other words, the necessary condition on DSWs $q_z$ is
\begin{equation}
    \max\limits_{\xi} q_z = q_{z'}^o > \sqrt{\varepsilon}.\label{eq:max_qz_condition}
\end{equation}
The range of possible values $q_z$ for DSWs is shown on the right in Fig.~\ref{fig:flat_surface} with $q_{z'}^o$ and $q_{z'}^e$.
After algebraic transformations one can find that Eq. \eqref{eq:max_qz_condition} reduces to
\begin{equation}
    0 < \xi < \frac{1}{2} - \frac{1}{\eta} \Leftrightarrow \varepsilon_\perp < \varepsilon < \frac{\varepsilon_\| - \varepsilon_\perp}{2}.\label{eq:true_dswm_condition}
\end{equation}
It immediately follows that anisotropy should be strong $\eta > 2$ for the existence of $\xi \in (0,1)$ and $\varepsilon$ accordingly.
It is clear that $q_{z'}^o, q_{z'}^e \xrightarrow{} \sqrt{\varepsilon_\perp}$ as $\varepsilon \xrightarrow{} \varepsilon_\|$, or equivalently $\xi \xrightarrow{} 1$, and $q_{z'}^o, q_{z'}^e \xrightarrow{} 0$ as $\varepsilon \xrightarrow{} \varepsilon_\perp$ corresponded to $\xi \xrightarrow{} 0$.
The dependence of $q_{z'}^o$ on $\xi$ is non-monotonic if $\eta > 2$.
There is a maximum $q_m = q_{z'}^o(\xi)$ at $\xi = \xi_m$.
After some algebraic transformations one can obtain analytical expression for it:
\begin{gather}
    q_m^2 = \varepsilon_\perp \frac{2 \tau^3}{\eta}, \quad \xi_m = \frac{(\tau-1)(1+2\tau)}{\eta(1+\tau)},\label{eq:qmxim}
\end{gather}
where $3\tau^2 = 1+\eta$.
The corresponding permittivity of isotropic medium is $\varepsilon_m = \varepsilon_\perp (1 + \eta\xi_m)$.
Thus, choosing the partnering materials for waveguide with $\varepsilon$ close to $\varepsilon_m$ theoretically allows one to get more guided DSWMs in Case B [Fig.~\ref{fig:sketch}(b)].
$\eta = 4, \xi = 1/4$ for the example under consideration [Fig.~\ref{fig:flat_surface}] which corresponds to the border case in Eq. \eqref{eq:qmxim}.
Thereby, all DSWMs are leaky in this approximation as said above.

\subsection{Simplified dispersion equation}
As mentioned before, losses are determined by $\operatorname{Im} q_{z,\nu}$ which are related to the radiation leakage in our case.
In waveguides with $R \gg R_0$ losses for DSWMs are small and tend to zero as $R \xrightarrow{} \infty$.
We can assume that $\operatorname{Im} q_{z,\nu} \ll \operatorname{Re} q_{z,\nu}$.
Therefore it is reasonable to find small imaginary correction $\operatorname{Im} q_{z,\nu}$ using the perturbation theory for equation \eqref{eq:main}.
Initial $q_{z',\nu}$ is defined so that $(q_{y',\nu}, q_{z',\nu})$ belongs to the DSWs dispersion curve.
Both of values $q_{z',\nu}$ and $q_{y',\nu}$ are significantly different from $0$ if $\xi$ is not close to $0$ or $1$.
This condition makes it possible to simplify Eq. \eqref{eq:main} with respect to $\nu \xrightarrow{} \infty$ that corresponds also to $R/R_0 \xrightarrow{} \infty$.
A similar procedure is used to estimate the Q--factor of whispering gallery modes \cite{Oraevsky2002Whispering-galleryWaves}.

Let us consider Case A as an example.
Using the asymptotic expansions for Bessel functions \cite{Abramowitz1965HandbookTables} and after tedious algebraic transformations, Eq. \eqref{eq:main} can be reduced to
\begin{equation}
    (\varepsilon_\| - \varepsilon)\varkappa_3(\varkappa_3 - \varkappa_1) = (\varepsilon_\perp \varkappa_2 + \varepsilon \varkappa_3)(\varkappa_1 + \varkappa_2) + i X_0.\label{eq:modified_equation}
\end{equation}
where $\varkappa_{1,2,3}$ is defined by
\begin{align*}
    q_z^2 + q_{y,\nu}^2 - \varkappa_1^2 &= \varepsilon, \\
    q_z^2 + q_{y,\nu}^2 - \varkappa_3^2 &= \varepsilon_\perp, \\
    \frac{q_{y,\nu}^2 - \varkappa_2^2}{\varepsilon_\|} + \frac{q_z^2}{\varepsilon_\perp} &= 1,
\end{align*}
and $q_{y,\nu} = \nu/R$.
The expression for $X_0$ is too cumbersome [See Appendix \ref{appendix:modified_equation}].
Here $\varkappa_1, \varkappa_2, \varkappa_3$ are the same as $k_1, k_2, k_3$ from Ref. \cite{Dyakonov1988} if we let $q$ and $\varphi$ be defined by $q \cos \varphi = q_z$ and $q \sin \varphi = \nu/R$.
Details of derivation Eq. \eqref{eq:modified_equation} are given in Appendix \ref{appendix:modified_equation}.
Term $X_0$ defines $\operatorname{Im} q_z$ and does not change $\operatorname{Re} q_z$ in the first-order perturbation.
$X_0 \in \mathbb{Re}$ has the right sign that $\operatorname{Im} q_z > 0$.
If waveguide radius $R \xrightarrow{} \infty$ then $X_0 \xrightarrow{} 0$ which is consistent with the above.
If we put $X_0 = 0$ and multiply Eq. \eqref{eq:modified_equation} by $(\varkappa_1+\varkappa_3)$ then Eq. \eqref{eq:modified_equation} coincides with the dispersion equation (9) in \cite{Dyakonov1988}.
The only difference is, $q_{y'}$ takes discrete values in our case.
One should solve Eq. \eqref{eq:modified_equation} for $q_z = \operatorname{Re} q_z + i \operatorname{Im} q_z$.
Equation \eqref{eq:modified_equation} works well only for $R$ large enough and greater than $R_0$ predicted by \eqref{eq:r0_weak}.
We attribute this to over-simplification expansions for large $R$, but it can be useful for some practical cases of the bent interface.

\begin{figure*}
    \includegraphics[width=17.8cm,clip,page=1]{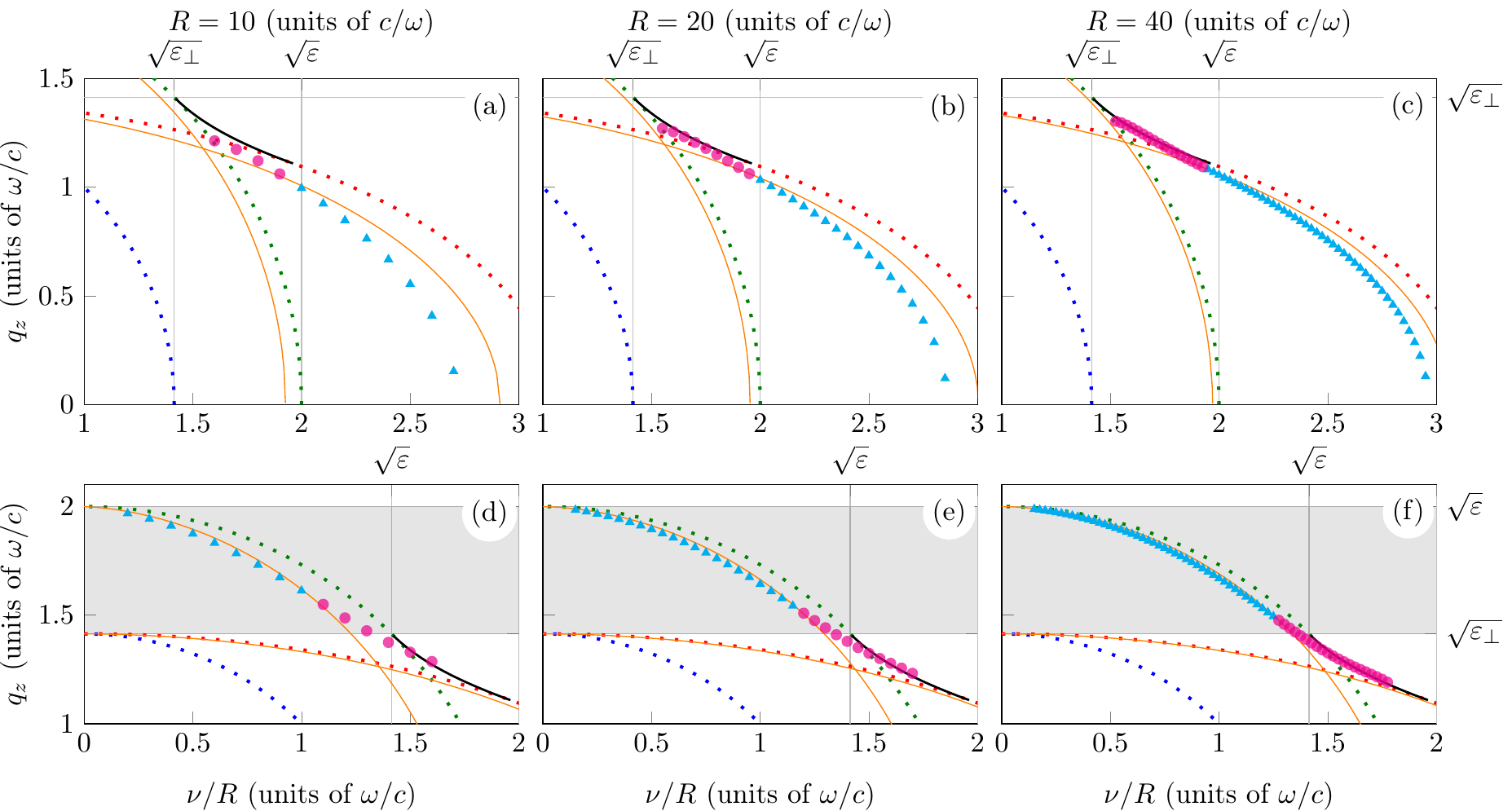}
    \caption{Generalized eigenmodes $(q_{z,\nu}, \nu/R)$ in anisotropic cylindrical waveguides of radius $R$ with frequency $\omega$. For these eigenmodes $\operatorname{Im}q_{z,\nu}=0$. Parameters of media: isotropic medium $\varepsilon = 4$ and uniaxial medium $\varepsilon_\| = 10, \varepsilon_\perp = 2$. The optical axis of the uniaxial medium coincides with the waveguide axis. (a)--(c) correspond to the waveguides from Case A [Fig.~\ref{fig:sketch}(a)]. (d)--(f) correspond to the waveguides from Case B [Fig.~\ref{fig:sketch}(b)]. Different columns correspond to waveguides of the same radius $R=10,20,40 \cdot c/\omega$. The green dotted curve shows the dispersion of TE- and TM-waves in the isotropic medium. The red and blue dotted curves show the dispersion of ordinary and extraordinary waves in the anisotropic medium as in Fig.~\ref{fig:flat_surface}. The black solid line shows the angular dispersion of Dyakonov surface waves at the flat interface. The orange curves show the boundaries of the region for surface modes in waveguide. Surface modes, also called Dyakonov-like surface waveguide mode, are marked with purple circles. Not surface modes are marked with cyan triangles. The gray region shows the region of guided modes in Case B [Fig.~\ref{fig:sketch}(b)].}
    \label{fig:dyakonov_waves_waveguide_modes}
\end{figure*}

\section{Eigenmodes in a arbitrary radius $R$ waveguides}\label{sec:arbitrary-waveguide}
If the radius $R$ of the waveguide is the same order as wavelength, i.e. in the isotropic medium, the previous approach could not be applied.
Then one should solve the exact equation \eqref{eq:main} numerically.
At first, we are looking for generalized eigenmodes which have $\operatorname{Im} q_{z,\nu} = 0$.
Let us consider waveguides with radii $R = 10, 20, 40$ as an example.
The full set of solutions $(\nu/R, q_{z,\nu})$ are rich in these cases.
Therefore, we have chosen a part of this set that will help for understanding.
The triangle and circle marks in Fig.~\ref{fig:dyakonov_waves_waveguide_modes} represent  the solutions $(\nu/R, q_{z,\nu})$ of the Eq. \eqref{eq:main}.
Figures \ref{fig:dyakonov_waves_waveguide_modes}(a)--\ref{fig:dyakonov_waves_waveguide_modes}(c) correspond to Case A.
Figures \ref{fig:dyakonov_waves_waveguide_modes}(d)--\ref{fig:dyakonov_waves_waveguide_modes}(f) correspond to Case B.
Figures \ref{fig:dyakonov_waves_waveguide_modes}(a) and \ref{fig:dyakonov_waves_waveguide_modes}(c), \ref{fig:dyakonov_waves_waveguide_modes}(b) and \ref{fig:dyakonov_waves_waveguide_modes}(d), \ref{fig:dyakonov_waves_waveguide_modes}(c) and \ref{fig:dyakonov_waves_waveguide_modes}(f) correspond to the different waveguide radius $R = 10, 20, 40$.
The dotted curves in figures correspond to the dispersion of plane waves, as in Fig.~\ref{fig:flat_surface}.
The orange curves in figures separate the cyan triangles from the magenta circles.
See Supplemental Material at [URL will be inserted by publisher] for more detailed plots, including a larger number of the solutions $(\nu/R, q_{z,\nu})$.

The modes marked with the magenta circles tend to the DSWs dispersion curve as $R$ increases in both cases A and B.
These modes are the exact DSWMs.
The field intensity distribution of these modes has only one maximum located at the surface of the waveguide.
For modes marked with cyan triangles, the maximum intensity is located inside the waveguide, as like as for WGMs in Case A or WMs in Case B.

\subsection{Surface waves definition}
In Case A all modes are leaky unlike Case B where guided modes exist for $q_z > \sqrt{\varepsilon_\perp}$.
The problem here is how to define surface waves regardless of whatever they are leaky or guided.
It is well known that Bessel functions $J_\nu(z), Y_\nu(z), H^{(1)}_\nu(z)$ change their behaviour at $z \approx \nu$.
Let us look at the region are limited by inequalities $k_o R < \nu$, $k_e R < \nu$ and $k R < \nu$.
If we denote $\nu/R = q_y$ then equations $k_o = q_y$, $k_e = q_y$, $k = q_y$ can be transformed to isofrequency contours $\Omega(q_z, q_y) = 0$ (angular dispersion curves) for plane waves in media.
Inequalities correspond to the outside area and thus the area of surface waves at the flat interface.
Therefore, we can call the modes $(\nu/R, q_{z,\nu})$ of the waveguide that falls into this area as surface modes.

This definition limits the region of surface waves too tightly and can discard modes that are sufficiently strong localized at the interface of the waveguide.
For example, using the previous definition one should discard all DSWMs in Fig.~\ref{fig:dyakonov_waves_waveguide_modes}(a) and almost all in Fig.~\ref{fig:dyakonov_waves_waveguide_modes}(d).
We suggest using a more precise definition based on the Bessel functions proprieties too.
A mode in the waveguide is surface if the following conditions are met:
(1) The field and intensity distribution inside waveguide should not have nodes and antinodes.
(2) The maximum of intensity distribution are located at $\rho = R$ exactly.
(3) If the fields outside is $E_z(\rho), H_z(\rho) \propto Y_\nu(k\rho)$ then $k R < x_0$ where $x_0$ is a value such that $\forall x < x_0: |Y_\nu(x)| > |Y_\nu(x_0)|$.
Thereby, $x_0$ is the solution to the $Y_\nu(x_0) = -Y_\nu(y'_{\nu,1})$, where $y'_{\nu,1}$ is the position of the first local maximum of $Y_\nu(y)$.
This definition slightly expands the area for surface modes introduced in the previous definition.
One can obtain analytic formulae for the boundary of the region using the asymptotic expansion of zeros of Bessel functions for high orders $\nu$.
The orange curves show this boundary in Fig.~\ref{fig:dyakonov_waves_waveguide_modes}(a)--\ref{fig:dyakonov_waves_waveguide_modes}(f).

\subsection{Waveguide eigenmodes (Case A)}
Let us analyze the eigenmodes of the waveguide in Case A.
Figures \ref{fig:dyakonov_waves_waveguide_modes}(a)--\ref{fig:dyakonov_waves_waveguide_modes}(c) show a part of modes in the waveguide for different radius $R$.
These modes have the largest value of $q_z$ at fixed $\nu$.
Modes with $q_z \approx 0$ are slowly propagating WGM-like modes.
The definition of surface waves region in the limit $\nu \xrightarrow{} \infty$ gives the following expressions for its boundaries:
\begin{align*}
    \left[q_y + h_o(q_y)\right]^2 + q_z^2 &> \varepsilon,\\
    \frac{\left[q_y + h_i(q_y)\right]^2}{\varepsilon_\|} + \frac{q_z^2}{\varepsilon_\perp} &> 1,
\end{align*}
where
\begin{equation}\label{eq:deltao-deltai}
\begin{aligned}
h_o(q_y) &\sim 0.275 R^{-2/3} q_y^{1/3},\\
h_i(q_y) &\sim 0.808 R^{-2/3} q_y^{1/3}
\end{aligned}
\end{equation}
and $q_y = \nu / R$.
The factor before $q_y^{1/3}$ in $h_{o,i}$ is $\propto R^{-2/3}$ and tends to $0$ as $R \xrightarrow{} \infty$.
It means that the region tends to the region defined by the first definition as for the region of surface waves on a flat interface.
The extended definition agrees with the simple one in this limit.
There is four DSWMs with $\nu = 16, 17, 18, 19$ in the waveguide with radius $R=10$ [Fig.~\ref{fig:dyakonov_waves_waveguide_modes}(a)].
The total number of DSWMs (9 for $R=20$, 17 for $R=40$) is increasing with increasing $R$ [Figs. \ref{fig:dyakonov_waves_waveguide_modes}(b)--\ref{fig:dyakonov_waves_waveguide_modes}(c)].

Tracing the dependence of $q_z$ on $\nu/R$ one could notice that DSWMs are going to be like WGM-like modes for $\nu > 20$.
It can be argued in this case that DSWMs come from WGMs due to the strong anisotropy of the medium inside the waveguide.
Unlike DSWMs, the conventional WGMs are localized inside the waveguide and near the interface.
The strong hybridization of E- and H-wave makes it possible to localize the WGM-like modes with smaller $\nu$ at the interface of the waveguide transforming it into DSWMs.

\subsection{Waveguide eigenmodes (Case B)}
In contrast to Case A, the existence of guided modes is possible in the region $q_z^2 > \varepsilon_\perp$.
This region is colored in gray in Figs. \ref{fig:dyakonov_waves_waveguide_modes}(d)--\ref{fig:dyakonov_waves_waveguide_modes}(f).
As in Case A, Figs. \ref{fig:dyakonov_waves_waveguide_modes}(d)--\ref{fig:dyakonov_waves_waveguide_modes}(f) show a part of the eigenmodes with the largest $q_z$ for different radius $R$.
Modes in the gray region are fundamental WMs and do not have radiation losses.
The second definition of the region of surface waves in the limit $\nu \xrightarrow{} \infty$ yields:
\begin{align*}
    \left[q_y + h_i(q_y)\right]^2 + q_z^2 &> \varepsilon,\\
    \frac{\left[q_y + h_o(q_y)\right]^2}{\varepsilon_\|} + \frac{q_z^2}{\varepsilon_\perp} &> 1,
\end{align*}
where $h_{o,i}$ is defined by Eq. \eqref{eq:deltao-deltai} and $q_y = \nu / R$.
The orange curves in Figs. \ref{fig:dyakonov_waves_waveguide_modes}(d)--\ref{fig:dyakonov_waves_waveguide_modes}(f) show the boundary of this region.
As $R \xrightarrow{} \infty$ the same situation happens, the extended region tends to be the region obtained by using the simple definition.

We obtain, by extended definition, six DSWMs with $\nu = 11\div16$ in the waveguide with radius $R=10$.
Unlike Case A, some of DSWMs with lower $\nu$ fall into the gray region of guided modes.
As $R$ increases [Figs. \ref{fig:dyakonov_waves_waveguide_modes}(e)--\ref{fig:dyakonov_waves_waveguide_modes}(f)], guided DSWMs fall into the gray region less and less deeply and become more like DSWs on a flat interface.
Tracing the dependence of maximal $q_z$ on $\nu/R$ for DSWMs in the same way one could notice that DSWMs are going to be like WMs as $\nu$ decreases.
Thus, the strong hybridization between E- and H-modes for larger $\nu$ makes it possible to localize WM exactly at the waveguide interface transforming it into DSWM in this case.

One can note the general properties of the DSWMs in both cases, such as quantization of $q_y$ for DSWs in cylindrical waveguide, the limited range of possible azimuthal numbers $\nu$ dependent on $\omega$ or radius $R$, the finite number of DSWMs as a consequence of previous property, the appearance of frequency dispersion for a fixed $\nu$ DSWM, the existence of the minimal radius $R$ of the waveguide or the minimal $\omega$ for which DSWMs exist.

\subsection{Field distribution}
\begin{figure}
    \centering
    \includegraphics[width=8.55cm,clip,page=5]{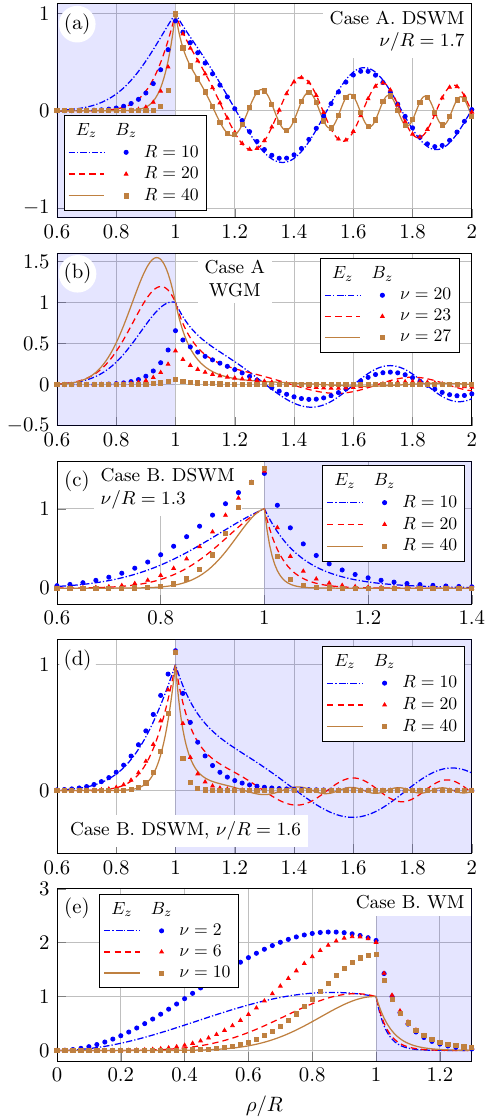}
    \caption{Plot of $E_z$, $B_z$ [arb. units] distributions for different types of modes in anisotropic waveguides. All distributions are normalized such as $E_z(R)=1$. The colored region shows the anisotropic medium. Case A: (a) DSWMs for different radius $R$ at fixed $\nu/R = 1.7$. (b) WGM-like modes for $R=10$ and different azimuthal number $\nu$. Case B: (c) guided DSWMs for different radius $R$ at fixed $\nu/R = 1.3$. (d) leaky DSWMs for different radius $R$ at fixed $\nu/R = 1.6$. (e) WMs for $R=10$ and different azimuthal number $\nu$.}
    \label{fig:fields}
\end{figure}

Figure \ref{fig:fields} shows distributions of $E_z$ and $B_z$ for different types of eigenmodes in waveguides under consideration.
Figures \ref{fig:fields}(a), \ref{fig:fields}(c), and \ref{fig:fields}(d) show field distributions in DSWMs.
Figure \ref{fig:fields}(b) shows field distribution in WGM-like modes.
Figure \ref{fig:fields}(e) shows field distribution in WMs.
The colored region in figures denotes the anisotropic medium.
Comparing the field distributions for different $R$ for a fixed $\nu/R$ [Figs. \ref{fig:fields}(a), \ref{fig:fields}(c), and \ref{fig:fields}(d)], we could conclude that the localization of DSWMs gets stronger as $R$ gets larger.
Another note is, only for DSWMs the ratio between $E_z$ and $H_z$ is close to 1.
It means that the hybridization of E- and H-wave in DSWM is significant as well as in the case of DSW.
Slow WGMs [Fig.~\ref{fig:fields}(b)] in Case A and WMs [Fig.~\ref{fig:fields}(e)] in Case B show the different behaviour.
E-wave dominates over H-wave in the WGM-like modes.
And in the WMs, H-wave dominates over E-wave vice versa.
This is due to the fact that E- and H-wave could be separated in Eq. \eqref{eq:main} when $q_z=0$ or $\nu=0$.
The maximum of $E_z$ for WGM-like mode is located inside the anisotropic medium and shifts further from the boundary as $\nu$ increases [Fig.~\ref{fig:fields}(b)].
It indicates that WGM-like modes are not truly surface.
Figures \ref{fig:fields}(c) and \ref{fig:fields}(d) show the difference between guided and leaky DSWMs in Case B [Figs. \ref{fig:sketch}(b), \ref{fig:dyakonov_waves_waveguide_modes}(d)--\ref{fig:dyakonov_waves_waveguide_modes}(f)].
For $\nu/R=1.3$ DSWMs are guided [Fig.~\ref{fig:fields}(c)] and for $\nu/R=1.7$ DSWMs are leaky [Fig.~\ref{fig:fields}(d)].
The guided DSWMs are quite similar to the WMs [Fig.~\ref{fig:fields}(e)] except that the maximum of $B_z$ for the second is located inside core not at the waveguide interface.
It is clear that for leaky DSWMs in Case B [Fig.~\ref{fig:fields}(d)] E-wave dominates over H-wave for $\rho > R$.
This hints to us that the radiation losses, in this case, will be associated mainly with E-wave tailor at $\rho \xrightarrow{} \infty$.

\section{Losses of DSWMs}\label{sec:losses}
Here we are not taking into account media losses, so $\operatorname{Im}\varepsilon = \operatorname{Im}\varepsilon_\perp=\operatorname{Im}\varepsilon_\| = 0$.
Therefore, losses of modes, in this case, are radiation losses due to the bending of the boundary of a cylindrical waveguide.
One of the useful characteristic parameters of this is figure-of-merit $\operatorname{FOM}$.
It shows how far mode is propagating along the waveguide in terms of the period along the waveguide axis:
\begin{equation}
    \operatorname{FOM} = \frac{\operatorname{Re} q_z}{\operatorname{Im} q_z}.
\end{equation}

In order to determine $\operatorname{Im} q_z$ one should solve \eqref{eq:main} choosing a diverging cylindrical wave $\propto H_\nu^{(1)}$ in the medium outside instead of standing wave $\propto Y_\nu$.
Real $q_z$ obtained in Fig. \ref{fig:dyakonov_waves_waveguide_modes} could be used as a starting point in the equation \eqref{eq:main} in order to calculate complex-valued $q_z$.
For well-defined modes such as WGMs, WMs, and DSWMs $\operatorname{Im}q_z$ is quite small.

\begin{figure}
    \centering
    \includegraphics[width=8.55cm,clip,page=1]{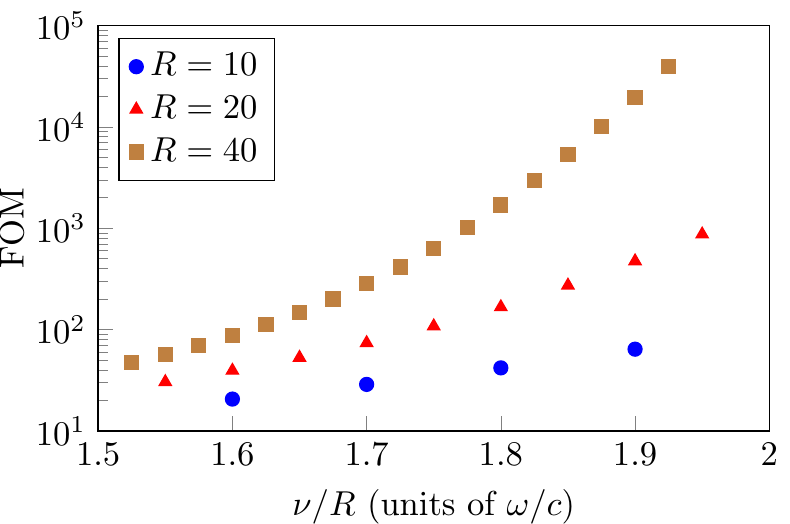}
    \caption{Figure-of-merit $\operatorname{FOM}$ of DSWMs for different waveguide radius $R$ in Case A [Fig.~\ref{fig:sketch}(a)]. For larger $R$ more DSWMs exist.}
    \label{fig:fom_dswm}
\end{figure}

We have analyzed Case A [Fig.~\ref{fig:sketch}(a)] only for the sake of simplicity.
It seems to us more attractive for experimental investigation.
Figure \ref{fig:fom_dswm} shows the comparison of $\operatorname{FOM}$ between DSWMs with different $\nu/R$ for different waveguide radius $R=10,20,40$.
For smaller $R$ DSWMs have smaller $\operatorname{FOM}$ that is quite evident.
Smaller $R$ leads to larger curvature of the boundary that follows larger radiation losses and smaller propagation length or $\operatorname{FOM}$ in other words.
Dyakonov-like surface waveguide modes that are close to WGM-like modes (larger $\nu/R$) have larger $\operatorname{FOM}$ and propagation length, and it is significantly growing faster for larger $R$.

\begin{figure}
    \centering
    \includegraphics[width=8.55cm,clip,page=2]{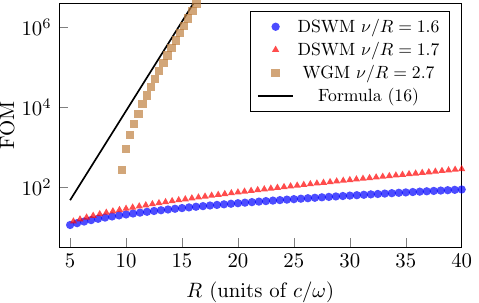}
    \caption{Comparison of figure-of-merit $\operatorname{FOM}$ dependence on waveguide radius $R$ between two DSWMs and WGM-like mode in Case A [Fig.~\ref{fig:sketch}(a)].}
    \label{fig:fom_dependence_on_r}
\end{figure}

Figure \ref{fig:fom_dependence_on_r} shows the difference in behaviour of $\operatorname{FOM}$ on $R$ for WGM-like mode with fixed $\nu/R=2.7$ and two DSWMs with fixed $\nu/R=1.6,1.7$.
$\operatorname{FOM}$ for WGM-like mode grows faster with radius $R$ than for DSWMs.
It means that WGM-like mode propagates much further than DSWMs.
The dependence of $\operatorname{FOM}$ on $R$ for selected DSWMs is quite weak and is not exponential in contrast to the WGM-like mode.
For WGM-like mode $\operatorname{FOM}$ could be analytically approximated in the limit $R \xrightarrow{} \infty$ using the fact that $\operatorname{Im}q_z \propto \Gamma \ll 1$ (See Appendix \ref{appendix:modified_equation}).
Let put $q_y=\nu/R$.
For fundamental WGM-like mode, the largest possible value of $q_z$ is limited by the dispersion of extraordinary waves \eqref{eq:extraordinary-wave}.
Thus, $q_z \approx \sqrt{\varepsilon_\perp(1-q_y^2/\varepsilon_\|)}$.
Then
\begin{equation}
    \operatorname{FOM}(R) \propto \Gamma^{-1} \underset{R \xrightarrow{} \infty}{\propto} \exp[\gamma(q_y) R],\label{eq:fom_wgm}
\end{equation}
where
\begin{equation}
    \gamma(q_y) = \frac{4}{3} \left[\psi\left(\frac{\varepsilon-q_z^2}{q_y^2}\right)\right]^{3/2},
\end{equation}
and $\psi(z)$ is defined by Eq. \eqref{eq:psi_def}.
It gives $\gamma = 1.04$ for WGM-like mode $q_y=2.7$.
Black dashed line in Fig.~\ref{fig:fom_dependence_on_r} shows the fitted dependence \eqref{eq:fom_wgm} with constant factor.
It perfectly fits the $\operatorname{FOM}$ calculated numerically for ${R>13}$.
For DSWMs we have not found any useful analytical expression for $\operatorname{FOM}$ on $R$.
This requires further research.

\section{Conclusion}\label{sec:conclusion}
The propagation of Dyakonov surface waves at the curved interface between isotropic and anisotropic uniaxial media has been theoretically investigated on the example of a cylindrical waveguide.
Dyakonov waves generally are not robust to interface bending and transform into leaky modes of the waveguide which have radiation losses.
We refer to these modes as Dyakonov-like surface waveguide modes (DSWMs).
These modes are well defined and have a distinctive feature:
It is strongly localized at the surface of the waveguide in contrast to whispering gallery modes and waveguide modes which are mostly localized inside the waveguide.
Losses are small in proportion to the ratio of the wavelength to the waveguide radius and tend to zero as radius tends to infinity.

Dyakonov waves behave differently depending on the side the interface bends to.
In the case of \textquotedblleft{}positive\textquotedblright{} curvature Dyakonov waves always have losses due to lack of guided modes in the waveguide.
Otherwise, in the case of \textquotedblleft{}negative\textquotedblright{} curvature it has been shown that some of DSWMs could be guided for specific material parameters if uniaxial is strongly anisotropic.
The conditions on permittivities have been determined under which the largest number of guided DSWMs could be obtained.
In these cases, the propagation lengths tend to infinity and are determined only by the losses in media.
This significantly distinguishes DSWM from surface plasmons-polariton propagating along plasmonic wires, which always have material absorption.

In another limit of small radii of curvature, DSWMs are no longer localized at the interface and disappear.
It has been shown that there is a minimum waveguide radius for which DSWMs exist.
Regardless of the type of waveguide, DSWMs have several common properties.
Waveguide structure leads to the quantization of one component of the wavevector and, therefore, the quantization of Dyakonov waves.
Because of this, DSWMs at fixed azimuthal number have a strong frequency dispersion in contrast to Dyakonov waves at the flat interface.
We have shown that whispering gallery modes in anisotropic cylinder continuously transforms to DSWMs as wavevector along waveguide $q_z$ increases and that conventional waveguide modes in isotropic medium placed inside uniaxial medium continuously transform to DSWMs as $q_z$ decreases.
This is due to the strong hybridization of E- and H-wave, as in the case of ordinary and extraordinary waves hybridization in Dyakonov surface waves at a flat interface.
We have shown that field distributions for DSWMs confirm it.

Thus, a new look at the anisotropic cylindrical waveguides supporting surface modes could be useful in applications for optics and photonics.

\begin{acknowledgements}
We would like to thank N.~S.~Averkiev for helpful discussions.
This work was supported by Russian Foundation for Basic Research (Grant No. 17-02-01234 A).
G.~K.~Yu. personally acknowledges the support by the Government of the Russian Federation (contract No. 14.W03.31.0011 at the Ioffe Institute of RAS).
\end{acknowledgements}

%


\appendix
\section{Field components}\label{appendix:fields}
In polar coordinates solutions of Eqs. \eqref{eq:ez} and \eqref{eq:hz} can be written as $E_z = A f_\nu(\rho) e^{i\nu\varphi}$ and $H_z = B g_\nu(\rho)e^{i\nu\varphi}$.
Then the other components of the electromagnetic field in cylindrical coordinates are expressed as follows
\begin{align}
	E_\rho &= \frac{i}{k_o^2}\left(A q_z \frac{\mathrm{d}f_\nu}{\mathrm{d}\rho} + B\frac{i\nu\omega}{c\rho} g_\nu\right)e^{i\nu\varphi}, \label{eq:erho}\\
	E_\varphi &= \frac{i}{k_o^2}\left(A\frac{i\nu q_z}{\rho} f_\nu - B\frac{\omega}{c}\frac{\mathrm{d}g_\nu}{\mathrm{d}\rho}\right)e^{i\nu\varphi}, \label{eq:ephi}\\
	H_\rho &= \frac{i}{k_o^2}\left(B q_z \frac{\mathrm{d}g_\nu}{\mathrm{d}\rho} - A\frac{i\nu\omega\varepsilon_\perp}{c\rho} f_\nu\right)e^{i\nu\varphi}, \label{eq:hrho}\\
	H_\varphi &= \frac{i}{k_o^2}\left(B \frac{i \nu q_z}{\rho} g_\nu + A\frac{\omega\varepsilon_\perp}{c}\frac{\mathrm{d}f_\nu}{\mathrm{d}\rho}\right)e^{i\nu\varphi}, \label{eq:hphi}\\
	D_{\rho,\varphi} &= \varepsilon_\perp E_{\rho,\varphi},\\
	D_z &= \varepsilon_\| E_z.
\end{align}
Constants $A$ and $B$ are independent in case when $q_z = 0$ or $\nu = 0$.
The first corresponds to whispering gallery modes ($\nu \gg 1$) running around in the waveguide cut and not propagating waveguide.
The latter corresponds to independent TE$_{0,1}$, TM$_{0,1}$ modes.
For all other cases, the ratio $A$ and $B$ is determined by boundary conditions.

\section{Dispersion equation}\label{appendix:main_equation}
Let us consider the general case of a waveguide when both media are uniaxial.
Let $R$ to be the radius of the waveguide.
Let both optical axes to be parallel to the $z$-axis and the waveguide axis.
The dielectric permittivity of the medium inside waveguide ${\hat{\varepsilon}^{\text{in}} = \operatorname{diag}(\varepsilon_\perp^{\text{in}},\varepsilon_\perp^{\text{in}},\varepsilon_\|^{\text{in}})}$ and outside ${\hat{\varepsilon}^{\text{out}} = \operatorname{diag}(\varepsilon_\perp^{\text{out}},\varepsilon_\perp^{\text{out}},\varepsilon_\|^{\text{out}})}$.
In what follows we use superscripts $^\text{in}$ and $^\text{out}$ for quantities in medium inside and medium outside respectively.
Let use the same notation introduced in Appendix \ref{appendix:fields}, Eq. \eqref{eq:ez} and Eq. \eqref{eq:hz}.
Then, $g_\nu(\rho)$ and $k_o^2 = \varepsilon_\perp \omega^2 /c^2 - q_z^2$ corresponds to H-wave and $f_\nu(\rho)$ and $k_e^2 = \varepsilon_\| k_o^2 / \varepsilon_\perp$ corresponds to E-wave.
Try to choose $f_\nu(\rho)$ and $g_\nu(\rho)$ in such a form that $f_\nu(R)=g_\nu(R)=1$:
\begin{gather}
    f^{\text{in}}_\nu(\rho) = \begin{cases}
    \dfrac{J_\nu(k_e^{\text{in}} \rho)}{J_\nu(k_e^{\text{in}} R)},&{k_e^{\text{in}}}^2>0,\\
    \dfrac{I_\nu(\varkappa_e^{\text{in}} \rho)}{I_\nu(\varkappa_e^{\text{in}} R)},&{\varkappa_e^{\text{in}}}^2 = -{k_e^{\text{in}}}^2>0,
    \end{cases} \\
    g^{\text{in}}_\nu(\rho) = \begin{cases}
    \dfrac{J_\nu(k_o^{\text{in}} \rho)}{J_\nu(k_o^{\text{in}} R)},&{k_o^{\text{in}}}^2>0,\\
    \dfrac{I_\nu(\varkappa_o^{\text{in}} \rho)}{I_\nu(\varkappa_o^{\text{in}} R)},&{\varkappa_o^{\text{in}}}^2 = -{k_o^{\text{in}}}^2>0,
    \end{cases}
\end{gather}
To search for eigenmodes and generalized eigenmodes ($\omega, q_z \in \mathbb{R}$) one should take functions outside as
\begin{gather}
    f^{\text{out}}_\nu(\rho) = \begin{cases}
    \dfrac{Y_\nu(k_e^{\text{out}} \rho)}{Y_\nu(k_e^{\text{out}} R)},&{k_e^{\text{out}}}^2>0,\\
    \dfrac{K_\nu(\varkappa_e^{\text{out}} \rho)}{K_\nu(\varkappa_e^{\text{out}} R)},&{\varkappa_e^{\text{out}}}^2 = -{k_e^{\text{out}}}^2>0,
    \end{cases} \\
    g^{\text{out}}_\nu(\rho) = \begin{cases}
    \dfrac{Y_\nu(k_o^{\text{out}} \rho)}{Y_\nu(k_o^{\text{out}} R)},&{k_o^{\text{out}}}^2>0,\\
    \dfrac{K_\nu(\varkappa_o^{\text{out}} \rho)}{K_\nu(\varkappa_o^{\text{out}} R)},&{\varkappa_o^{\text{out}}}^2 = -{k_o^{\text{out}}}^2>0.
    \end{cases}
\end{gather}
In order to find radiation losses for leaky modes ($\omega \in \mathbb{R}$ and $q_z \in \mathbb{C}$) Bessel function $Y_\nu$ should be replaced by $H^{(1)}_\nu$ instead.
Here, $J_\nu(z), Y_\nu(z)$ are Bessel function of the First and the Second kind, $I_\nu(z), K_\nu(z)$ are modified Bessel function of the First and the Second kind, $H^{(1)}_\nu(z)$ is Hankel function of the First kind.

Chosen form of $f_\nu$ and $g_\nu$ immediately satisfies the boundary conditions for $E_z$ and $B_z$ on the waveguide surface.
Thus, the constants $A$ and $B$ remain undefined [Eqs. \eqref{eq:erho}--\eqref{eq:hphi}]
The relation between them is determined by other boundary conditions, for example the continuity of $E_\varphi$ and $H_\varphi$.
Boundary conditions can be rewritten as $M \big(A\ B\big)^{\text{T}} = 0$, where $M$ is 2x2 matrix.
$\det M = 0$ leads to the dispersion equation
\begin{widetext}
\begin{equation}\label{eq:main_equation_general}
    \left(\frac{{g^{\text{in}}_\nu}'(R)}{{k^{\text{in}}_o}^2} - \frac{{g^{\text{out}}_\nu}'(R)}{{k^{\text{out}}_o}^2}\right) \left(\frac{\varepsilon^{\text{in}}_\perp{f^{\text{in}}_\nu}'(R)}{{k^{\text{in}}_o}^2} - \frac{\varepsilon^{\text{out}}_\perp{f^{\text{out}}_\nu}'(R)}{{k^{\text{out}}_o}^2}\right)
    = \frac{\nu^2 q_z^2 \omega^2}{{k^{\text{out}}_o}^4 {k^{\text{in}}_o}^4 c^2 R^2} \left(\varepsilon^{\text{out}}_\perp - \varepsilon^{\text{in}}_\perp\right)^2,
\end{equation}
\end{widetext}
where $f' \equiv \mathrm{d}f/\mathrm{d}\rho$.
In Case A (anisotropic medium for $\rho < R$ and isotropic medium for $\rho > R$)
dielectric permittivities $\varepsilon^\text{out}_\perp = \varepsilon^\text{out}_\| = \varepsilon$ and ${k_o^\text{out}}^2 = {k_e^\text{out}}^2 = k^2$.
Let $F_{e,h}^{\text{in},\text{out}}$ to be defined as
\begin{alignat}{4}
    &F_h^\text{in} &&= \frac{\mathrm{d}g_\nu^\text{in}(\rho)}{\mathrm{d}\rho}\bigg|_{\rho=R}, \quad &&F_e^\text{in} &&= \frac{\mathrm{d}f_\nu^\text{in}(\rho)}{\mathrm{d}\rho}\bigg|_{\rho=R}, \\
    &F_h^\text{out} &&= \frac{\mathrm{d}g_\nu^\text{out}(\rho)}{\mathrm{d}\rho}\bigg|_{\rho=R}, \quad &&F_e^\text{out} &&= \frac{\mathrm{d}f_\nu^\text{out}(\rho)}{\mathrm{d}\rho}\bigg|_{\rho=R}.
\end{alignat}
Then, Eq. \eqref{eq:main_equation_general} can be transformed into Eq. \eqref{eq:main}.
For modes defined by the dispersion equation \eqref{eq:main_equation_general} the ratio between $A$ and $B$ is given by expression
\begin{equation}
    \frac{A}{B} = -\,\frac{i R c (k_o^{\text{out}} k_o^{\text{in}})^2}{\nu q_z \omega (\varepsilon - \varepsilon_\perp)}\left(\frac{{g^{\text{in}}_\nu}'(R)}{{k_o^{\text{in}}}^2} - \frac{{g^{\text{out}}_\nu}'(R)}{{k_o^{\text{out}}}^2}\right).
\end{equation}

\section{Minimal waveguide radius $R_0$}\label{appendix:minimal_radius}
We are define $R_0$ as the minimal waveguide radius $R$ from which at least one DSWM can be exist.
This is necessary but not sufficient condition of DSWMs existence.
$R_0$ depends on the range $(q_{y'}^o, q_{y'}^e)$ cause $q_{y'} \approx \nu/R_0$.
Using expressions for $\sin^2\vartheta_1$ and $\sin^2\vartheta_2$ from original Dyakonov's work \cite{Dyakonov1988} one can get:
\begin{align}
    \left(q_{y}^o\right)^2 &= \frac{4\pi^2}{\lambda_\varepsilon^2}\frac{\xi}{2}\left(1 - \eta\xi + \sqrt{(1-\eta\xi)^2 + 4\eta}\right), \label{eq:appendix_qy1}\\
    \left(q_{y}^e\right)^2 &= \frac{4\pi^2}{\lambda_\varepsilon^2}\frac{\xi (1+\eta)^3}{(1+\eta\xi)^2(1+2\eta-\eta\xi)}, \label{eq:appendix_qy2}
\end{align}
where $\lambda_\varepsilon$ is wavelength in isotropic medium.
If $1/R \leq q_{y}^e - q_{y}^o$ then at least one DSWM could exist.
$R_0$ is defined by equality $1/R_0 = q_{y}^e - q_{y}^o$.
Expanding Equations \eqref{eq:appendix_qy1} and \eqref{eq:appendix_qy2} for $\eta$ in the case of weak anisotropy $\eta \ll 1$ up to the third order gives
\begin{equation}
    \frac{1}{R_0} = \frac{\lambda_\varepsilon}{2\pi}\,\eta^2 \sqrt{\xi} (1-\xi)^2,\label{eq:app-r0}
\end{equation}
which is equivalent to Eq. \eqref{eq:r0_weak}.
Expression \eqref{eq:app-r0} works well until $\eta < 1/3$.
It is not easy to obtain simple closed-form expression of $1/R_0$ for the case of strong anisotropy $\eta \gg 1$ due to strong dependence of $q_y^o$ and $q_y^e$ at ${\xi \sim \eta^{-1/2}}$.
Therefore, we give a semi-empirical expression
\begin{equation}
    \frac{2\pi R_0}{\lambda_\varepsilon} = \frac{1+\eta \xi}{(\sqrt{1+\eta^2}-1)\sqrt{\xi}(1-\xi)^2},\label{eq:r0_strong}
\end{equation}
which works well for $1 < \eta \lesssim 10$ covering most practical cases.

\section{Modified DSWs dispersion equation}\label{appendix:modified_equation}
As it said above $\nu \xrightarrow{} \infty$ with $R/\lambda_\varepsilon \xrightarrow{} \infty$ for DSWMs because $q_y \approx \nu/R$ and $q_y$ is the same order as $q_z$ if $\xi$ does not close to 1 or 0.
Under such conditions when ${R/\lambda_\varepsilon \xrightarrow{} \infty}$ equation \eqref{eq:main} can be simplified by using high order $\nu$ expansions for Bessel functions \cite{Abramowitz1965HandbookTables}:
\begin{gather}
    \frac{J'_\nu(\nu z)}{J_\nu(\nu z)} \sim \frac{\sqrt{1-z^2}}{z} \left(-\frac{\operatorname{Ai}'(\psi)}{\sqrt{\psi}\operatorname{Ai}(\psi)}\right),\label{eq:besselj_expansion}\\
    \frac{{H^{(1)}_\nu}'(\nu z)}{{H^{(1)}_\nu}(\nu z)} \sim \frac{\sqrt{1-z^2}}{z} \left(-\frac{1}{\sqrt{\psi}}\frac{\operatorname{Ai}'(\psi)-i\operatorname{Bi}'(\psi)}{\operatorname{Ai}(\psi)-i\operatorname{Bi}(\psi)}\right),\label{eq:hankelh_expansion}
\end{gather}
where
\begin{equation}
    \psi(z) = \left[\frac{3\nu}{2}\left(\ln\frac{1+\sqrt{1-z^2}}{z}-\sqrt{1-z^2}\right)\right]^{2/3},\label{eq:psi_def}
\end{equation}
and $z < 1$, $\operatorname{Ai}(\psi)$ and $\operatorname{Bi}(\psi)$ are Airy functions,
These expansions work very well until $z \lesssim 0.9$.

Let us consider the case of anisotropic waveguide [Fig.~\ref{fig:sketch}(a)].
Then, argument is $z = k_o R / \nu$ or $z = k_e R / \nu$ in Eq. \eqref{eq:besselj_expansion} and $z = k R / \nu$ in Eq. \eqref{eq:hankelh_expansion}.
Regions of $(q_z, \nu/R)$ where expansions \eqref{eq:besselj_expansion} and \eqref{eq:hankelh_expansion} can be used are defined by
\begin{align}
    \frac{k_o R}{\nu} &< 1 - \delta & &\Rightarrow & q_z^2 + \frac{\nu^2}{R^2}(1-\delta)^2 &> \varepsilon_\perp, \\
    \frac{k_e R}{\nu} &< 1 - \delta & &\Rightarrow & \frac{q_z^2}{\varepsilon_\perp} + \frac{\nu^2}{R^2 \varepsilon_\|}(1-\delta)^2 &> 1, \\
    \frac{k R}{\nu} &< 1 - \delta & &\Rightarrow & q_z^2 + \frac{\nu^2}{R^2}(1-\delta)^2 &> \varepsilon,
\end{align}
where $\delta \approx 0.1$.
In the case $\delta = 0$ it corresponds to the region for surface waves (DSWs in our case) at the flat interface.
For $\delta > 0$ the region is slightly shrunken because expansions Eqs. \eqref{eq:besselj_expansion} and \eqref{eq:hankelh_expansion} do not work well for $z=1$ or equivalently $\delta = 0$.

Imaginary part of Eq. \eqref{eq:hankelh_expansion} only determines radiation losses as small imaginary correction to $q_z$.
Terms in brackets in Eqs. \eqref{eq:besselj_expansion} and \eqref{eq:hankelh_expansion} tend to 1 as $\nu \xrightarrow{} \infty$.
This is due to $\psi \xrightarrow{} \infty$ as $\nu \xrightarrow{} \infty$ if $1 - z \gtrsim \nu^{-2/3}$ which is satisfied for $z \gtrsim 0.9$ and $\nu > 1$.
Let us neglect real corrections to $q_z$ and let imaginary first-order correction term $\Gamma$ be defined by
\begin{equation}
    \Gamma = \operatorname{Im}\left(-\frac{1}{\sqrt{\psi}}\frac{\mathrm{Ai}'(\psi) - i\mathrm{Bi}'(\psi)}{\mathrm{Ai}(\psi) - i\mathrm{Bi}(\psi)}\right),\label{eq:gamma_loss}
\end{equation}
where $\psi = \psi(kR/\nu)$ [Eq. \eqref{eq:psi_def}].
$\Gamma$ is a small parameter, $\Gamma \approx e^{-4\psi^{3/2}/3} \ll 1$ as $\nu \xrightarrow{} \infty$.
Let us consider $\Gamma$ as perturbation parameter in Eq. \eqref{eq:main}.
After tedious algebraic transformations one can obtain the simpler equation \eqref{eq:modified_equation} in the first-order perturbation
\begin{equation}
    (\varepsilon_\| - \varepsilon)\varkappa_3(\varkappa_3 - \varkappa_1) = (\varepsilon_\perp \varkappa_2 + \varepsilon \varkappa_3)(\varkappa_1 + \varkappa_2) + iX_0,\label{eq:modified_equation_app}
\end{equation}
where
\begin{align}
    q_z^2 + q_y^2 - \varkappa_1^2 &= \varepsilon, \\
    q_z^2 + q_y^2 - \varkappa_3^2 &= \varepsilon_\perp, \\
    \frac{q_y^2 - \varkappa_2^2}{\varepsilon_\|} + \frac{q_z^2}{\varepsilon_\perp} &= 1,
\end{align}
\begin{equation}
    X_0 = \Gamma \eta \frac{k_o^2 \varkappa_1 (k^2 (\varepsilon_\perp \varkappa_2 + \varepsilon \varkappa_1) + 2\varepsilon k_o^2 \varkappa_1)}{k^2 (\varkappa_1 + \varkappa_3) (\varkappa_2 - \varkappa_3)},
\end{equation}
and $q_y = \nu / R$.
The equation \eqref{eq:modified_equation_app} [Eq. \eqref{eq:modified_equation}] is solved for $q_z$ for a fixed $\nu$ and $R$.
It is clear that $X_0 \xrightarrow{} 0$ as $\Gamma \xrightarrow{} 0$ or $\nu \xrightarrow{} \infty$, equivalently.
If we put $X_0 = 0$ and redefine $q_z = q \cos \vartheta$ and $q_y = q \sin \vartheta$ then Eq. \eqref{eq:modified_equation_app} will become the dispersion equation for Dyakonov waves at the flat interface from Ref. \cite{Dyakonov1988}.
Taking into account $X_0$ leads to a small correction $\operatorname{Im} q_z \propto \Gamma$ which defines the radiation losses of DSWMs in the cylindrical waveguide.

The described procedure can be applied to the other case of isotropic waveguide inside anisotropic medium [Fig.~\ref{fig:sketch}(b)].
The difference will be in the definition $X_0$ which is mainly determined in this case by the radiation of cylindrical E-wave in the anisotropic medium.


\bibliography{literature}

\begin{thebibliography}{18}%
\makeatletter
\providecommand \@ifxundefined [1]{%
 \@ifx{#1\undefined}
}%
\providecommand \@ifnum [1]{%
 \ifnum #1\expandafter \@firstoftwo
 \else \expandafter \@secondoftwo
 \fi
}%
\providecommand \@ifx [1]{%
 \ifx #1\expandafter \@firstoftwo
 \else \expandafter \@secondoftwo
 \fi
}%
\providecommand \natexlab [1]{#1}%
\providecommand \enquote  [1]{``#1''}%
\providecommand \bibnamefont  [1]{#1}%
\providecommand \bibfnamefont [1]{#1}%
\providecommand \citenamefont [1]{#1}%
\providecommand \href@noop [0]{\@secondoftwo}%
\providecommand \href [0]{\begingroup \@sanitize@url \@href}%
\providecommand \@href[1]{\@@startlink{#1}\@@href}%
\providecommand \@@href[1]{\endgroup#1\@@endlink}%
\providecommand \@sanitize@url [0]{\catcode `\\12\catcode `\$12\catcode
  `\&12\catcode `\#12\catcode `\^12\catcode `\_12\catcode `\%12\relax}%
\providecommand \@@startlink[1]{}%
\providecommand \@@endlink[0]{}%
\providecommand \url  [0]{\begingroup\@sanitize@url \@url }%
\providecommand \@url [1]{\endgroup\@href {#1}{\urlprefix }}%
\providecommand \urlprefix  [0]{URL }%
\providecommand \Eprint [0]{\href }%
\providecommand \doibase [0]{https://doi.org/}%
\providecommand \selectlanguage [0]{\@gobble}%
\providecommand \bibinfo  [0]{\@secondoftwo}%
\providecommand \bibfield  [0]{\@secondoftwo}%
\providecommand \translation [1]{[#1]}%
\providecommand \BibitemOpen [0]{}%
\providecommand \bibitemStop [0]{}%
\providecommand \bibitemNoStop [0]{.\EOS\space}%
\providecommand \EOS [0]{\spacefactor3000\relax}%
\providecommand \BibitemShut  [1]{\csname bibitem#1\endcsname}%
\let\auto@bib@innerbib\@empty
\bibitem [{\citenamefont {Takayama}\ \emph {et~al.}(2008)\citenamefont
  {Takayama}, \citenamefont {Crasovan}, \citenamefont {Johansen}, \citenamefont
  {Mihalache}, \citenamefont {Artigas},\ and\ \citenamefont
  {Torner}}]{Takayama2008DyakonovReview}%
  \BibitemOpen
  \bibfield  {author} {\bibinfo {author} {\bibfnamefont {O.}~\bibnamefont
  {Takayama}}, \bibinfo {author} {\bibfnamefont {L.-C.}\ \bibnamefont
  {Crasovan}}, \bibinfo {author} {\bibfnamefont {S.~K.}\ \bibnamefont
  {Johansen}}, \bibinfo {author} {\bibfnamefont {D.}~\bibnamefont {Mihalache}},
  \bibinfo {author} {\bibfnamefont {D.}~\bibnamefont {Artigas}},\ and\ \bibinfo
  {author} {\bibfnamefont {L.}~\bibnamefont {Torner}},\ }\href
  {https://doi.org/10.1080/02726340801921403} {\bibfield  {journal} {\bibinfo
  {journal} {Electromagnetics}\ }\textbf {\bibinfo {volume} {28}},\ \bibinfo
  {pages} {126} (\bibinfo {year} {2008})}\BibitemShut {NoStop}%
\bibitem [{\citenamefont {Takayama}\ \emph {et~al.}(2017)\citenamefont
  {Takayama}, \citenamefont {Bogdanov},\ and\ \citenamefont
  {Lavrinenko}}]{Takayama2017PhotonicInterfaces}%
  \BibitemOpen
  \bibfield  {author} {\bibinfo {author} {\bibfnamefont {O.}~\bibnamefont
  {Takayama}}, \bibinfo {author} {\bibfnamefont {A.~A.}\ \bibnamefont
  {Bogdanov}},\ and\ \bibinfo {author} {\bibfnamefont {A.~V.}\ \bibnamefont
  {Lavrinenko}},\ }\href {https://doi.org/10.1088/1361-648x/aa8bdd} {\bibfield
  {journal} {\bibinfo  {journal} {Journal of Physics: Condensed Matter}\
  }\textbf {\bibinfo {volume} {29}},\ \bibinfo {pages} {463001} (\bibinfo
  {year} {2017})}\BibitemShut {NoStop}%
\bibitem [{\citenamefont {D'yakonov}(1988)}]{Dyakonov1988}%
  \BibitemOpen
  \bibfield  {author} {\bibinfo {author} {\bibfnamefont {M.}~\bibnamefont
  {D'yakonov}},\ }\href {http://www.jetp.ac.ru/cgi-bin/dn/e_067_04_0714.pdf}
  {\bibfield  {journal} {\bibinfo  {journal} {Sov. Phys. JETP}\ }\textbf
  {\bibinfo {volume} {67}},\ \bibinfo {pages} {714} (\bibinfo {year}
  {1988})}\BibitemShut {NoStop}%
\bibitem [{\citenamefont {Averkiev}\ and\ \citenamefont
  {Dyakonov}(1990)}]{Averkiev1990}%
  \BibitemOpen
  \bibfield  {author} {\bibinfo {author} {\bibfnamefont {N.~S.}\ \bibnamefont
  {Averkiev}}\ and\ \bibinfo {author} {\bibfnamefont {M.}~\bibnamefont
  {Dyakonov}},\ }\href@noop {} {\bibfield  {journal} {\bibinfo  {journal} {Opt.
  Spectrosc.}\ }\textbf {\bibinfo {volume} {68}},\ \bibinfo {pages} {653}
  (\bibinfo {year} {1990})}\BibitemShut {NoStop}%
\bibitem [{\citenamefont {Walker}\ \emph {et~al.}(1998)\citenamefont {Walker},
  \citenamefont {Glytsis},\ and\ \citenamefont
  {Gaylord}}]{Walker1998SurfaceInterfaces}%
  \BibitemOpen
  \bibfield  {author} {\bibinfo {author} {\bibfnamefont {D.~B.}\ \bibnamefont
  {Walker}}, \bibinfo {author} {\bibfnamefont {E.~N.}\ \bibnamefont
  {Glytsis}},\ and\ \bibinfo {author} {\bibfnamefont {T.~K.}\ \bibnamefont
  {Gaylord}},\ }\href {https://doi.org/10.1364/JOSAA.15.000248} {\bibfield
  {journal} {\bibinfo  {journal} {J. Opt. Soc. Am. A}\ }\textbf {\bibinfo
  {volume} {15}},\ \bibinfo {pages} {248} (\bibinfo {year} {1998})}\BibitemShut
  {NoStop}%
\bibitem [{\citenamefont {Cojocaru}(2015)}]{Cojocaru2015DyakonovInterface}%
  \BibitemOpen
  \bibfield  {author} {\bibinfo {author} {\bibfnamefont {E.}~\bibnamefont
  {Cojocaru}},\ }\href {https://doi.org/10.1364/JOSAA.32.000782} {\bibfield
  {journal} {\bibinfo  {journal} {J. Opt. Soc. Am. A}\ }\textbf {\bibinfo
  {volume} {32}},\ \bibinfo {pages} {782} (\bibinfo {year} {2015})}\BibitemShut
  {NoStop}%
\bibitem [{\citenamefont {Jacob}\ and\ \citenamefont
  {Narimanov}(2008)}]{Jacob2008OpticalMetamaterials}%
  \BibitemOpen
  \bibfield  {author} {\bibinfo {author} {\bibfnamefont {Z.}~\bibnamefont
  {Jacob}}\ and\ \bibinfo {author} {\bibfnamefont {E.~E.}\ \bibnamefont
  {Narimanov}},\ }\href {https://doi.org/10.1063/1.3037208} {\bibfield
  {journal} {\bibinfo  {journal} {Applied Physics Letters}\ }\textbf {\bibinfo
  {volume} {93}},\ \bibinfo {pages} {221109} (\bibinfo {year}
  {2008})}\BibitemShut {NoStop}%
\bibitem [{\citenamefont {John A.~Polo}\ \emph {et~al.}(2007)\citenamefont
  {John A.~Polo}, \citenamefont {Nelatury},\ and\ \citenamefont
  {Lakhtakia}}]{JohnA.Polo2007SurfaceInterface}%
  \BibitemOpen
  \bibfield  {author} {\bibinfo {author} {\bibfnamefont {J.}~\bibnamefont {John
  A.~Polo}}, \bibinfo {author} {\bibfnamefont {S.~R.}\ \bibnamefont
  {Nelatury}},\ and\ \bibinfo {author} {\bibfnamefont {A.}~\bibnamefont
  {Lakhtakia}},\ }\href {https://doi.org/10.1364/JOSAA.24.002974} {\bibfield
  {journal} {\bibinfo  {journal} {J. Opt. Soc. Am. A}\ }\textbf {\bibinfo
  {volume} {24}},\ \bibinfo {pages} {2974} (\bibinfo {year}
  {2007})}\BibitemShut {NoStop}%
\bibitem [{\citenamefont {Agarwal}\ \emph {et~al.}(2009)\citenamefont
  {Agarwal}, \citenamefont {Jr},\ and\ \citenamefont
  {Lakhtakia}}]{Agarwal_2009}%
  \BibitemOpen
  \bibfield  {author} {\bibinfo {author} {\bibfnamefont {K.}~\bibnamefont
  {Agarwal}}, \bibinfo {author} {\bibfnamefont {J.~A.~P.}\ \bibnamefont {Jr}},\
  and\ \bibinfo {author} {\bibfnamefont {A.}~\bibnamefont {Lakhtakia}},\ }\href
  {https://doi.org/10.1088/1464-4258/11/7/074003} {\bibfield  {journal}
  {\bibinfo  {journal} {Journal of Optics A: Pure and Applied Optics}\ }\textbf
  {\bibinfo {volume} {11}},\ \bibinfo {pages} {74003} (\bibinfo {year}
  {2009})}\BibitemShut {NoStop}%
\bibitem [{\citenamefont {Gao}\ \emph {et~al.}(2010)\citenamefont {Gao},
  \citenamefont {Lakhtakia},\ and\ \citenamefont
  {Lei}}]{Gao2010Dyakonov-TammHandedness}%
  \BibitemOpen
  \bibfield  {author} {\bibinfo {author} {\bibfnamefont {J.}~\bibnamefont
  {Gao}}, \bibinfo {author} {\bibfnamefont {A.}~\bibnamefont {Lakhtakia}},\
  and\ \bibinfo {author} {\bibfnamefont {M.}~\bibnamefont {Lei}},\ }\href
  {https://doi.org/10.1103/PhysRevA.81.013801} {\bibfield  {journal} {\bibinfo
  {journal} {Phys. Rev. A}\ }\textbf {\bibinfo {volume} {81}},\ \bibinfo
  {pages} {13801} (\bibinfo {year} {2010})}\BibitemShut {NoStop}%
\bibitem [{\citenamefont {Kim}\ \emph {et~al.}(2006)\citenamefont {Kim},
  \citenamefont {Yang}, \citenamefont {Lee}, \citenamefont {Lee}, \citenamefont
  {Lee},\ and\ \citenamefont {{Woo-Jin-Jung}}}]{Kim2006LeakyWaveguide}%
  \BibitemOpen
  \bibfield  {author} {\bibinfo {author} {\bibfnamefont {W.-K.}\ \bibnamefont
  {Kim}}, \bibinfo {author} {\bibfnamefont {W.-S.}\ \bibnamefont {Yang}},
  \bibinfo {author} {\bibfnamefont {H.-M.}\ \bibnamefont {Lee}}, \bibinfo
  {author} {\bibfnamefont {H.-Y.}\ \bibnamefont {Lee}}, \bibinfo {author}
  {\bibfnamefont {M.-H.}\ \bibnamefont {Lee}},\ and\ \bibinfo {author}
  {\bibnamefont {{Woo-Jin-Jung}}},\ }\href
  {https://doi.org/10.1364/OE.14.013043} {\bibfield  {journal} {\bibinfo
  {journal} {Opt. Express}\ }\textbf {\bibinfo {volume} {14}},\ \bibinfo
  {pages} {13043} (\bibinfo {year} {2006})}\BibitemShut {NoStop}%
\bibitem [{\citenamefont {Kotelnikov}\ and\ \citenamefont
  {Stupakov}(2015)}]{Kotelnikov2015ElectromagneticCylinder}%
  \BibitemOpen
  \bibfield  {author} {\bibinfo {author} {\bibfnamefont {I.~A.}\ \bibnamefont
  {Kotelnikov}}\ and\ \bibinfo {author} {\bibfnamefont {G.~V.}\ \bibnamefont
  {Stupakov}},\ }\href
  {https://doi.org/https://doi.org/10.1016/j.physleta.2015.02.013} {\bibfield
  {journal} {\bibinfo  {journal} {Physics Letters A}\ }\textbf {\bibinfo
  {volume} {379}},\ \bibinfo {pages} {1187 } (\bibinfo {year}
  {2015})}\BibitemShut {NoStop}%
\bibitem [{\citenamefont {Fang}\ and\ \citenamefont
  {Sun}(2015)}]{Fang2015NanoplasmonicCircuits}%
  \BibitemOpen
  \bibfield  {author} {\bibinfo {author} {\bibfnamefont {Y.}~\bibnamefont
  {Fang}}\ and\ \bibinfo {author} {\bibfnamefont {M.}~\bibnamefont {Sun}},\
  }\href {https://doi.org/10.1038/lsa.2015.67} {\bibfield  {journal} {\bibinfo
  {journal} {Light: Science {\&} Applications}\ }\textbf {\bibinfo {volume}
  {4}},\ \bibinfo {pages} {e294} (\bibinfo {year} {2015})}\BibitemShut
  {NoStop}%
\bibitem [{\citenamefont {Snyder}\ and\ \citenamefont
  {Love}(1984)}]{Snyder1984OpticalTheory}%
  \BibitemOpen
  \bibfield  {author} {\bibinfo {author} {\bibfnamefont {A.~W.}\ \bibnamefont
  {Snyder}}\ and\ \bibinfo {author} {\bibfnamefont {J.~D.}\ \bibnamefont
  {Love}},\ }\href {https://doi.org/10.1007/978-1-4613-2813-1} {\emph {\bibinfo
  {title} {{Optical Waveguide Theory}}}}\ (\bibinfo  {publisher} {Springer
  US},\ \bibinfo {address} {Boston, MA},\ \bibinfo {year} {1984})\BibitemShut
  {NoStop}%
\bibitem [{\citenamefont {Kajorndejnukul}\ \emph {et~al.}(2019)\citenamefont
  {Kajorndejnukul}, \citenamefont {Artigas},\ and\ \citenamefont
  {Torner}}]{Kajorndejnukul2019ConformalMetamaterialsb}%
  \BibitemOpen
  \bibfield  {author} {\bibinfo {author} {\bibfnamefont {V.}~\bibnamefont
  {Kajorndejnukul}}, \bibinfo {author} {\bibfnamefont {D.}~\bibnamefont
  {Artigas}},\ and\ \bibinfo {author} {\bibfnamefont {L.}~\bibnamefont
  {Torner}},\ }\href {https://doi.org/10.1103/PhysRevB.100.195404} {\bibfield
  {journal} {\bibinfo  {journal} {Phys. Rev. B}\ }\textbf {\bibinfo {volume}
  {100}},\ \bibinfo {pages} {195404} (\bibinfo {year} {2019})}\BibitemShut
  {NoStop}%
\bibitem [{\citenamefont
  {Sumetsky}(2004)}]{Sumetsky2004Whispering-gallery-bottleEtalonb}%
  \BibitemOpen
  \bibfield  {author} {\bibinfo {author} {\bibfnamefont {M.}~\bibnamefont
  {Sumetsky}},\ }\href {https://doi.org/10.1364/OL.29.000008} {\bibfield
  {journal} {\bibinfo  {journal} {Opt. Lett.}\ }\textbf {\bibinfo {volume}
  {29}},\ \bibinfo {pages} {8} (\bibinfo {year} {2004})}\BibitemShut {NoStop}%
\bibitem [{\citenamefont
  {Oraevsky}(2002)}]{Oraevsky2002Whispering-galleryWaves}%
  \BibitemOpen
  \bibfield  {author} {\bibinfo {author} {\bibfnamefont {A.~N.}\ \bibnamefont
  {Oraevsky}},\ }\href {https://doi.org/10.1070/QE2002v032n05ABEH002205}
  {\bibfield  {journal} {\bibinfo  {journal} {Quantum Electronics}\ }\textbf
  {\bibinfo {volume} {32}},\ \bibinfo {pages} {377} (\bibinfo {year}
  {2002})}\BibitemShut {NoStop}%
\bibitem [{\citenamefont {Abramowitz}\ and\ \citenamefont
  {Stegun}(1965)}]{Abramowitz1965HandbookTables}%
  \BibitemOpen
  \bibfield  {author} {\bibinfo {author} {\bibfnamefont {M.}~\bibnamefont
  {Abramowitz}}\ and\ \bibinfo {author} {\bibfnamefont {I.~A.}\ \bibnamefont
  {Stegun}},\ }\href {https://dl.acm.org/citation.cfm?id=1098650} {\emph
  {\bibinfo {title} {{Handbook of mathematical functions, with formulas,
  graphs, and mathematical tables,}}}}\ (\bibinfo  {publisher} {Dover
  Publications},\ \bibinfo {year} {1965})\ p.\ \bibinfo {pages}
  {1046}\BibitemShut {NoStop}%
\end{thebibliography}%

\end{document}


\title{Supplemental material for "Dyakonov-like surface waves in anisotropic cylindrical waveguides"}

\author{K.~Yu. Golenitskii}
\email[]{golenitski.k@mail.ioffe.ru}
\affiliation{Ioffe Institute, 194021 St.~Petersburg, Russia}

\author{A.~A. Bogdanov}
\affiliation{Ioffe Institute, 194021 St.~Petersburg, Russia}
\affiliation{ITMO University, 197101 St.~Petersburg, Russia}

\date{\today}

\maketitle

\section*{Eigenmodes in anisotropic fibers}
\begin{figure*}
    \includegraphics[width=15.8cm,clip,page=2]{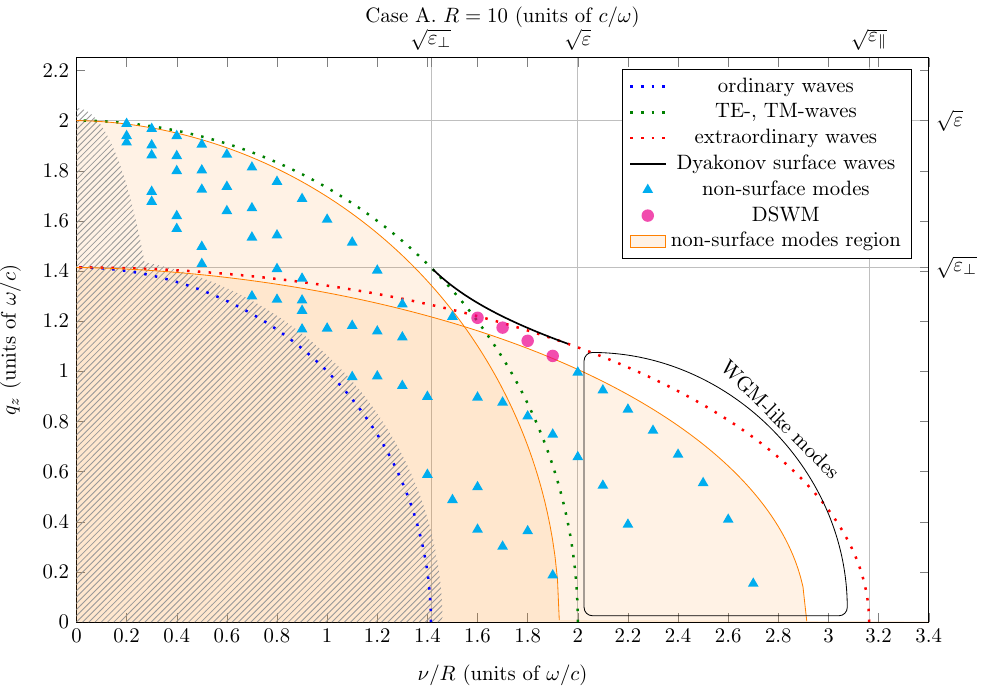}
    \caption{Plot of eigenmodes $(\nu/R, q_{z,\nu})$ in the anisotropic waveguide of radius $R = 10 c/\omega$. The core of the waveguide is considering uniaxial anisotropic with permittivity $\varepsilon_\perp = 2$, $\varepsilon_\| = 10$. The outside medium is isotropic dielectric $\varepsilon = 4$. All modes are leaky.}
    \label{fig:a10}
\end{figure*}

\begin{figure*}
    \includegraphics[width=15.8cm,clip,page=3]{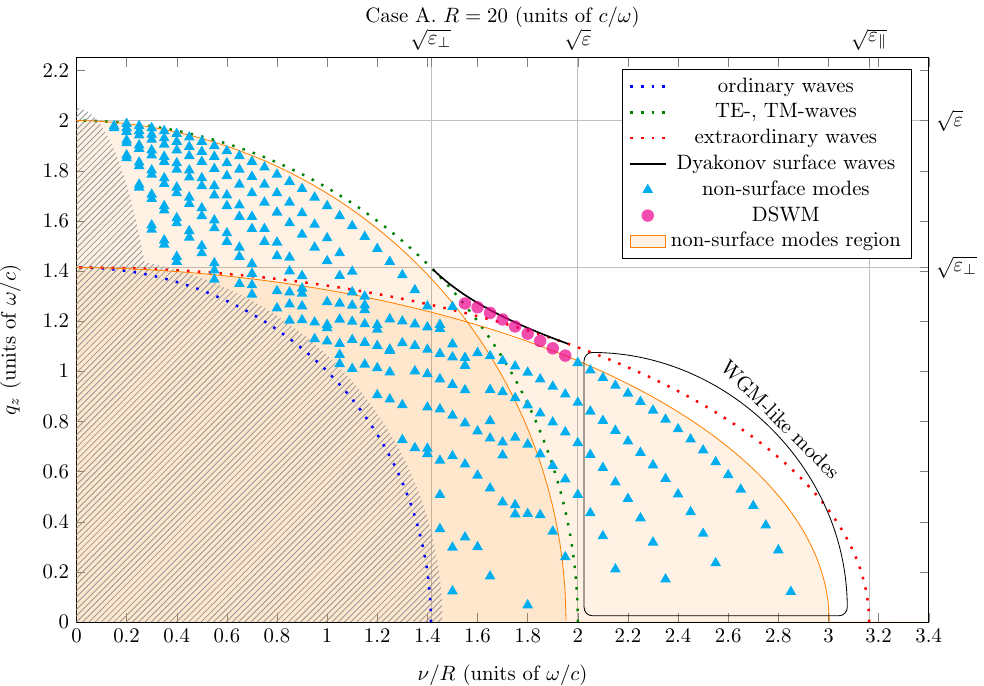}
    \caption{Plot of eigenmodes $(\nu/R, q_{z,\nu})$ in the anisotropic waveguide of radius $R = 20 c/\omega$. The core of the waveguide is considering uniaxial anisotropic with permittivity $\varepsilon_\perp = 2$, $\varepsilon_\| = 10$. The outside medium is isotropic dielectric $\varepsilon = 4$. All modes are leaky.}
    \label{fig:a20}
\end{figure*}

\begin{figure*}
    \includegraphics[width=15.8cm,clip,page=4]{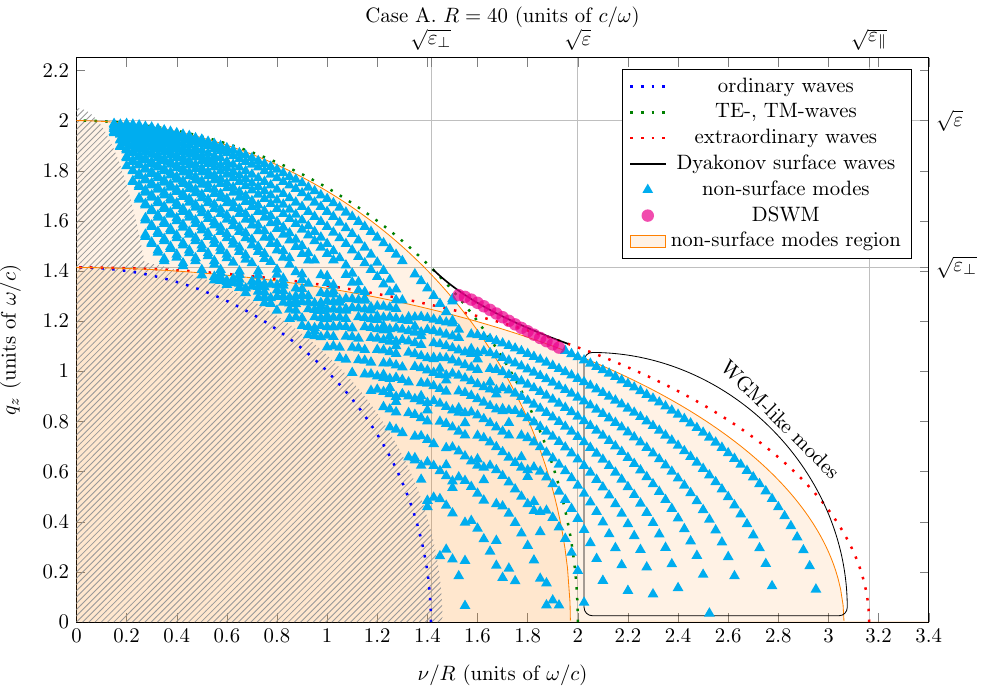}
    \caption{Plot of eigenmodes $(\nu/R, q_{z,\nu})$ in the anisotropic waveguide of radius $R = 40 c/\omega$. The core of the waveguide is considering uniaxial anisotropic with permittivity $\varepsilon_\perp = 2$, $\varepsilon_\| = 10$. The outside medium is isotropic dielectric $\varepsilon = 4$. All modes are leaky.}
    \label{fig:a40}
\end{figure*}

\begin{figure*}
    \includegraphics[width=15.8cm,clip,page=5]{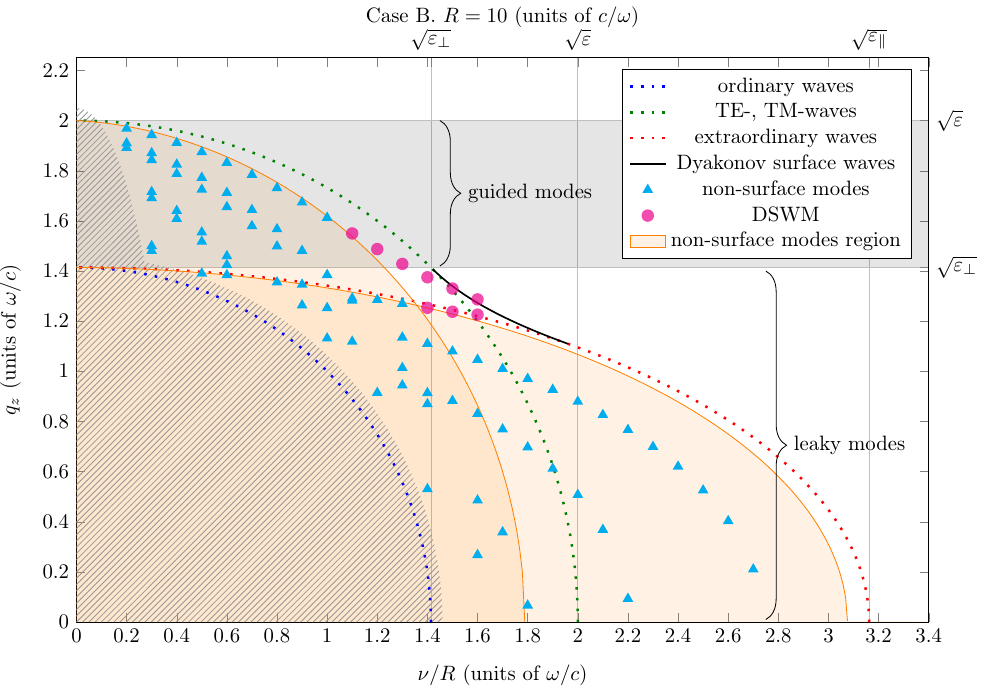}
    \caption{Plot of eigenmodes $(\nu/R, q_{z,\nu})$ in the anisotropic waveguide of radius $R = 10 c/\omega$. The core of the waveguide is considering isotropic dielectric $\varepsilon = 4$. The outside medium is uniaxial anisotropic medium with permittivity $\varepsilon_\perp = 2$, $\varepsilon_\| = 10$.}
    \label{fig:b10}
\end{figure*}

\begin{figure*}
    \includegraphics[width=15.8cm,clip,page=6]{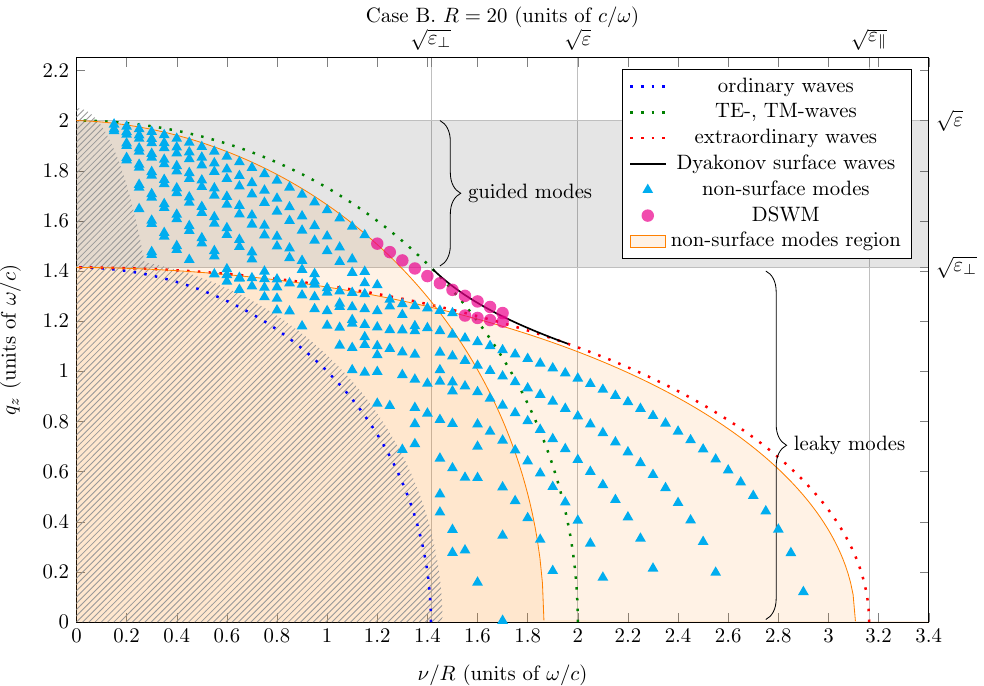}
    \caption{Plot of eigenmodes $(\nu/R, q_{z,\nu})$ in the anisotropic waveguide of radius $R = 20 c/\omega$. The core of the waveguide is considering isotropic dielectric $\varepsilon = 4$. The outside medium is uniaxial anisotropic medium with permittivity $\varepsilon_\perp = 2$, $\varepsilon_\| = 10$.}
    \label{fig:b20}
\end{figure*}

\begin{figure*}
    \includegraphics[width=15.8cm,clip,page=7]{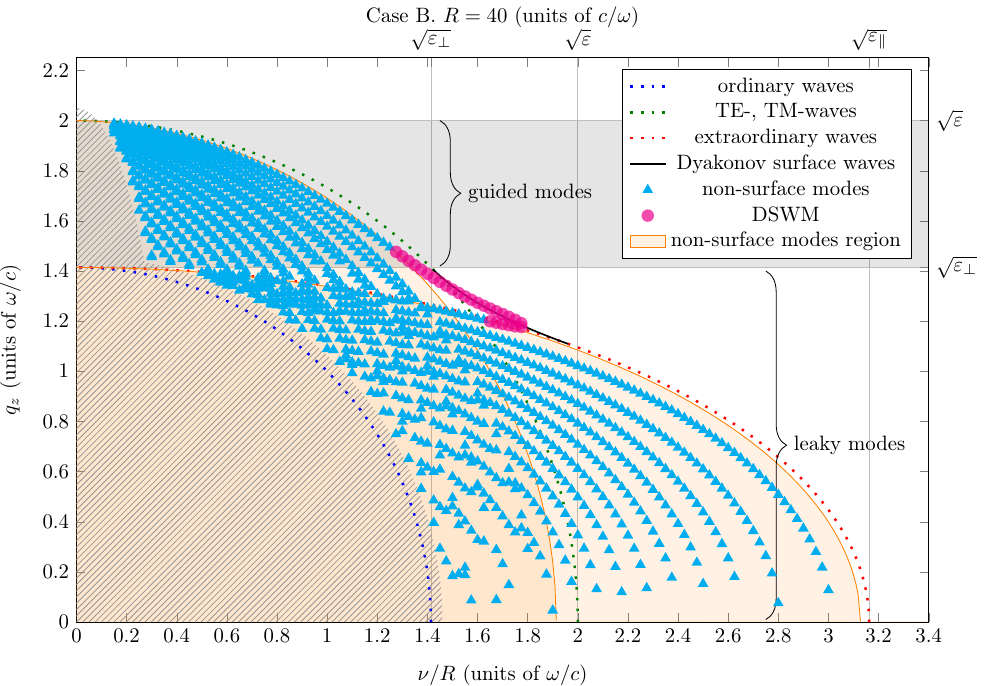}
    \caption{Plot of eigenmodes $(\nu/R, q_{z,\nu})$ in the anisotropic waveguide of radius $R = 40 c/\omega$. The core of the waveguide is considering isotropic dielectric $\varepsilon = 4$. The outside medium is uniaxial anisotropic medium with permittivity $\varepsilon_\perp = 2$, $\varepsilon_\| = 10$.}
    \label{fig:b40}
\end{figure*}

Complete set of solutions $(\nu/R, q_{z,\nu})$ of the dispersion equation for eigenmodes in cylindrical waveguides are rich.
The interface between uniaxial anisotropic medium with permittivity $\varepsilon_\perp = 2, \varepsilon_\| = 10$ and isotropic medium with permittivity $\varepsilon = 4$ is considered as an example.
Here we present larger sets of the solutions for 6 cases considered in the main text.
Figures \ref{fig:a10}--\ref{fig:a40} correspond to the waveguides with the anisotropic medium inside ($\rho < R$) and the isotropic medium outside ($\rho > R$) that we refer to as Case A.
Figures \ref{fig:b10}--\ref{fig:b40} correspond to the waveguides with the isotropic medium inside and the anisotropic medium outside that we refer to as Case B.
The hatched region on all figures shows the region where we have not looked for solutions.
It does not provide additional information to understanding.

There are two values $q_z$ at fixed $\nu/R$ for DSWM with the largest $\nu$ (magenta circles) in Figs. \ref{fig:b10}--\ref{fig:b40}.
It happens due to the definition of surface waves in a cylindrical waveguide.
The boundary of orange regions in Fig. \ref{fig:a10}--\ref{fig:b40} is defined in the limit $\nu \xrightarrow{} \infty$ only.
The definition of surface waves can fail exactly at the boundary if we consider finite $R$ and $\nu$. 
So we should manually check modes that are close to this boundary.

